\documentclass[aps,prd,reprint,floatfix,superscriptaddress,lengthcheck,sort&compress,letterpaper,nofootinbib]{revtex4-1}

\usepackage{graphicx}
\usepackage{dcolumn}
\usepackage{bm}
\usepackage{dsfont}
\usepackage{amsmath}
\usepackage[usenames]{color}
\usepackage{multirow}
\usepackage[colorlinks=true,linkcolor=blue,citecolor=blue,urlcolor=blue]{hyperref}
\usepackage[normalem]{ulem}
\usepackage{microtype}

\newcommand{\gS}[1]{#1\!\!\!\!\!\not~}

\newcommand{\pslash}{\gS{p}}

\newcommand{\ii}{\textrm{i}}
\newcommand{\beq}{\begin{equation}}
\newcommand{\eeq}{\end{equation}}
\newcommand{\bal}{\begin{align}}
\newcommand{\eal}{\end{align}}
\newcommand{\MeV}{\:\text{MeV}}
\newcommand{\GeV}{\:\text{GeV}}
\newcommand{\nc}{\ensuremath{\eta_c}}
\newcommand{\Jp}{\ensuremath{J/\psi}}

\begin{document}

\title{Four-quark states with charm quarks in a two-body Bethe-Salpeter approach}

\author{Nico Santowsky}
 \email[e-mail: ]{nico.santowsky@theo.physik.uni-giessen.de}
\affiliation{Institut f\"ur Theoretische Physik, Justus-Liebig Universit\"at Gie{\ss}en, 35392 Gie{\ss}en, Germany}
\affiliation{Helmholtz Forschungsakademie Hessen f\"ur FAIR (HFHF),
	GSI Helmholtzzentrum f\"ur Schwerionenforschung, Campus Gie{\ss}en, 35392 Gie{\ss}en, Germany}
\author{Christian S. Fischer}
 \email[e-mail: ]{christian.fischer@theo.physik.uni-giessen.de (corresponding author)}
\affiliation{Institut f\"ur Theoretische Physik, Justus-Liebig Universit\"at Gie{\ss}en, 35392 Gie{\ss}en, Germany}
\affiliation{Helmholtz Forschungsakademie Hessen f\"ur FAIR (HFHF),
	GSI Helmholtzzentrum f\"ur Schwerionenforschung, Campus Gie{\ss}en, 35392 Gie{\ss}en, Germany}

\date{\today}

\begin{abstract}

We study the internal structure of a range of four-quark states with charm quark contributions using a two-body 
Bethe-Salpeter equation. Thereby, we examine charmonium-like states with hidden charm and quark content 
$c\bar{c}q\bar{q}$, open-charm states with quark content $cc\bar{q}\bar{q}$ and all-charm states with 
$cc\bar{c}\bar{c}$. In particular we study the internal competition between meson-meson components and 
diquark-antidiquark components in the wave functions of these states. Our results indicate that the 
$\chi_{c1}(3872)$ and the $Z_c(3900)$ are predominantly $D\bar D^*$ states and that the recently discovered 
open-charm state $T_{cc}^+$ is dominated by an internal $DD^*$ component. In both cases the diquark components
are negligible. For the all-charm state $X(6900)$
with as yet unknown quantum numbers we identify candidates in the excitation spectra of $0^+$ and $1^+$ states. 
Furthermore, our framework serves to provide predictions for further, yet undiscovered open and hidden charm 
four-quark states.

\end{abstract}

\maketitle


\section{\label{sec:1}Introduction}
The discovery of the $\chi_{c1}$(3872) in 2003 by the Belle Collaboration \cite{Belle:2003nnu} is considered as 
the birth of exotic spectroscopy for states including heavy quarks $\{c,b\}$. Over the years, many more states 
were discovered in the mass region of charmonia and bottomonia, which cannot be explained by the conventional 
quark model, see e.g. \cite{Esposito:2016noz,Ali:2017jda,Brambilla:2019esw} for an overview. Whereas the quantum 
numbers of several exotic states (traditionally called $X$ and $Y$ states) are compatible with those from ordinary 
quarkonia, the so-called $Z$ states carry an electric charge, entailing a quark content of at least four quarks.
As a consequence, the notion of four-quark states became a paradigm for the discussion of all exotic heavy quark 
states exposed so far. Further support for this picture is obtained from the first candidate for an open charm
four-quark state, $T_{cc}^+$, with quark content $cc\bar{u}\bar{d}$ recently discovered by the LHCb-collaboration
\cite{LHCb:2021vvq} and the first all-charm state with quark content $cc\bar{c}\bar{c}$, 
the $X(6900)$, also discovered by LHCb \cite{LHCb:2020bwg}.

The internal structure of these exotic states is still heavily debated. States, like the $\chi_{c1}(3872)$, that 
are close to mesonic decay thresholds have been advocated as meson molecules with possible small admixtures 
of other components \cite{Guo:2017jvc}, but other interpretations are debated as well \cite{Esposito:2021vhu}. 
A clear distinction between a mesonic molecule and a compact four-quark state (e.g. built from diquark-antidiquark 
components) requires a detailed analysis of its associated line shape extracted from experiment. While first 
high quality results have been made available by the LHCb collaboration using fits to experimental data 
\cite{LHCb:2020xds}, final clarification of this matter may have to wait for direct measurements planned 
in the upcoming PANDA experiment \cite{PANDA:2018zjt,PANDA:2021ozp}. 

On the theory side, exotic candidates with heavy quarks have been described in a variety of approaches such 
as quark models \cite{Maiani:2004vq, Ebert:2007rn, Giron:2020wpx, Yang:2021hrb}, 
lattice QCD 
\cite{Prelovsek:2013cra,Ikeda:2013vwa, Prelovsek:2014swa, Padmanath:2015era,Francis:2016hui,Bicudo:2017szl,Cheung:2017tnt,Francis:2018jyb,Junnarkar:2018twb,Leskovec:2019ioa}, 
sum rules \cite{Albuquerque:2018jkn}, effective theories \cite{Wang:2013cya,Guo:2017jvc,Baru:2021ddn} 
or functional methods using a four-body Faddeev-Yakubovsky equation 
\cite{Wallbott:2019dng, Eichmann:2020oqt, Wallbott:2020jzh}.
While many of these studies investigate certain aspects of four-quark states, a full understanding seems 
only possible by taking multiple configurations into account simultaneously instead of assuming a certain 
internal structure a priori. Anticipating a clustering into internal two-quark states (provided by strong 
two-body forces), there are three different structures which may contribute to a four-quark state with quark 
content $c\bar{c}q\bar{q}$: 
(i)~a heavy-light meson-meson/molecular state where two heavy-light mesons interact with each other~\cite{Guo:2017jvc},
(ii)~a hadro-charmonium~\cite{Dubynskiy:2008mq} with a heavy $c\bar{c}$ core and a light $q\bar{q}$ pair surrounding it and
(iii)~a diquark-antidiquark state with strongly interacting heavy-light diquarks~\cite{Jaffe:2004ph}. 
In addition, other structures are possible. If the quantum numbers allow, there could be a sizeable
$c\bar{c}$-component. In some cases even meson three-body effects could play an important role 
\cite{LHCb:2021auc,Du:2021zzh}. Or, if the interaction between the (anti-)quarks is dominated by 
irreducible three- and four-body forces, a compact four-quark state may even arise with no 
preferred internal clustering. 

In this paper we study the internal structure of heavy-light and all-heavy four-quark states using a coupled 
system of covariant two-body Bethe-Salpeter equations that allow for a competition between different internal 
structures. This coupled system was firstly formulated in \cite{Heupel:2012ua} and recently extended to an 
investigation of resonances in the complex $P^2$ plane \cite{Santowsky:2021ugd}, where it yielded a 
qualitative description of the $f_0$ and $a_0$ states in light scalar nonet. It is derived from the generic 
four-body description studied in \cite{Wallbott:2019dng, Wallbott:2020jzh} and allows in principle also to study 
the effective mixing of four-quark states with quark-antiquark components \cite{Santowsky:2020pwd}. However, due 
to technical complexities involved, in this work we will stick to the pure four-quark picture and study the coupling to 
two-quark states in a future work. 

The paper is organized as follows: First, we briefly introduce the two-body Bethe-Salpeter equation in section
\ref{sec:2}. In section \ref{sec:3} technical details are discussed. Then, in section \ref{sec:4}, 
we present our results for charmonium-like candidates with quark content $c\bar{c}q\bar{q}$, open-charm ones 
with $cc\bar{q}\bar{q}$ and all-charm states with $cc\bar{c}\bar{c}$. In the end, we give a short summary and 
draw final conclusions.
\section{\label{sec:2}The four-quark two-body Bethe-Salpeter equation\label{thedsebseapproach}}
In order to make this article self-contained, we briefly repeat here the derivation of the (coupled set of)
effective two-body Bethe-Salpeter equations from the corresponding four-body equation of the four-quark state.
Additional details may be found in
\cite{Heupel:2012ua,Kvinikhidze:2014yqa,Santowsky:2020pwd,Kvinikhidze:2021kzu,Santowsky:2021ugd}.

The $2n$-quark Green's function $G^{(2n)}$ can be expressed via the $T$ matrix, the interacting part of the $S$ matrix,
\begin{equation}
	G^{(2n)} = G_0^{(n)} + G_0^{(n)} T^{(2n)} G_0^{(n)},
\end{equation}
where $G_0^{(n)}$ is the product of $n$ non-interacting, but fully dressed quark propagators. 
The $T$ matrix may be expressed via the $n$-quark scattering kernel $K^{(n)}$ as follows,
\begin{equation}
	T^{(2n)} = K^{(n)} + K^{(n)} G_0^{(n)} T^{(2n)}.\label{t-matrix}
\end{equation}
As bound states and resonances are poles of the $T$ matrix~\cite{ParticleDataGroup:2020ssz}, one obtains
\begin{equation}
	T^{(2n)}\xrightarrow{P^2\rightarrow-M^2}\frac{\Psi^{(n)}\bar\Psi^{(n)}}{P^2+M^2}
	\label{t-matrix-ansatz}
\end{equation}
in the proximity of the mass pole, where the Bethe-Salpeter amplitude $\Psi^{(n)}$ and its conjugate define 
the corresponding pole residue. Inserting this into (\ref{t-matrix}) and comparing the residues yields the 
homogeneous $n$-quark Bethe-Salpeter equation,
\begin{equation}
	\Psi^{(n)} = K^{(n)} G_0^{(n)} \Psi^{(n)}. \label{bse}
\end{equation}
In the case $n=4$ the kernel $K$ can be decomposed into irreducible two-, three- and four-quark correlations,
\begin{equation}
	K^{(4)} = \tilde{K}^{(2)} + \tilde{K}^{(3)} + \tilde{K}^{(4)}.
	\label{eq-2-3-4-body-forces}
\end{equation}
We neglect the three- and four-quark interaction kernels and set $\tilde{K}^{(3)}=\tilde{K}^{(4)}=0$. 
A priori, the justification for this approximation is on physics ground only\footnote{A similar approximation 
	was already applied successfully in the baryon sector where the diquark-quark picture led to a spectrum 
	in one-to-one agreement with experiment, see e.g. \cite{Eichmann:2016hgl, Eichmann:2016yit}. Whereas 
	in the baryon sector it can be shown explicitly that the leading part of the irreducible three-body 
	interaction (in terms of a skeleton expansion) is small \cite{Sanchis-Alepuz:2017jjd}, we don't have 
	such strict arguments concerning the four-body kernel that is relevant in this work.}. 
If we assume that the internal structure of four-quark states may be expressed in terms of meson-meson, 
hadro-charmonium or diquark-antidiquark components, then two-body forces must dominate over three- and 
four-body ones. Apart from lattice QCD, this assumption is inherent to all approaches to four-quark states 
known by us, and we shall adopt it here as well. 

The contribution $\tilde{K}^{(2)}$ containing all irreducible two- body interactions inside the four-quark state contains various incarnations of the two-body scattering kernel $K^{(2)}$ between two quarks $i$ and $j$:
\begin{align}
	\tilde{K}^{(2)}	&	=\underbrace{{K}^{(2)}_{12}S^{-1}_3S^{-1}_4+{K}^{(2)}_{34}S^{-1}_1S^{-1}_2
		-{K}^{(2)}_{12}{K}^{(2)}_{34}}_{=:\tilde{K}^{(2)}_{(12)(34)}}
	+ \text{perm.}\nonumber\\
	&   = \sum_a \tilde{K}^{(2)}_a \label{eq-2bodykernels}
\end{align}
Explicit indices $1,2,3,4$ denote the four (anti-)quarks as ingredients of the four-quark bound state and the summation over $a$ picks up the three possible combinations $(12)(34), (13)(24), (14)(23)$ of two-body interactions.

In order to be able to extract a two-body Bethe-Salpeter equation for the four-quark $q\bar{q}q\bar{q}$ state, we slightly reformulate the problem \cite{Heupel:2012ua}. First, we define a four-body $T$-matrix $T_a$ that is generated by $\tilde{K}^{(2)}_a$:
\begin{equation}
	T_a = \tilde{K}^{(2)}_a + \tilde{K}^{(2)}_a G_0^{(4)} T_a = \tilde{K}^{(2)}_a +  T_a G_0^{(4)} \tilde{K}^{(2)}_a\,. \label{Ta}
\end{equation}
Furthermore, we note that the BSA, Eq.~(\ref{bse}), can be split into three separate parts by inserting Eq.~(\ref{eq-2bodykernels})
\begin{align}
	\Psi = \sum_a \tilde{K}^{(2)}_a G_0^{(4)}\:\Psi := \sum_a \Psi_a\,.
\end{align}
Acting with $T_a G_0^{(4)}$ onto $\Psi$ and using (\ref{Ta}) one then obtains
\begin{equation}
	\Psi_a=T_a \,G_0^{(4)}\,(\Psi-\Psi_a)=\sum_{b\neq a}\,T_a\,G_0^{(4)}\,\Psi_b\,, \label{eq-masterbse}
\end{equation}
which is still an exact four-body equation apart from neglecting the kernels $\tilde{K}^{(3)}$ and $\tilde{K}^{(4)}$.

Since the $T$ matrices $T_a$ contain effects from two-body interactions in the same combination of (anti-)quark legs only, they are prone to develop singularities in the respective channels, with the quantum numbers of mesons and (anti-)diquarks. The two-body approximation of the four-body equation then amounts to replacing $T_a$ with a pole ansatz analogously to Eq.~(\ref{t-matrix-ansatz}). Assuming that the spin-momentum structure of the Bethe-Salpeter amplitudes factorizes, the full amplitude $\Psi$ can then be decomposed into meson-meson and diquark-antidiquark substructures $\Phi_{a}$. We thus obtain
\begin{equation}
	\Psi_a = \left(\Gamma_{12}\otimes\Gamma_{34}\right)\:G_0^{(2,2)}\:\Phi_{a}
	\label{eq-4bsastructure}
\end{equation}
for $a=(12)(34)$ and similar expressions for the other combinations. Here, $G_0^{(2,2)}$ is a combination of two meson propagators or a diquark and an antidiquark propagator, respectively, and $\Gamma_{ij}$ are the corresponding two-body Bethe-Salpeter amplitudes, i.e. $\Gamma:=\Psi^{(2)}$. The representation Eq.~(\ref{eq-4bsastructure}) is in some sense a `physical basis' in that it builds a representation of $\Psi_a$ in terms of reduced internal Dirac, flavour and colour structure from a physical picture. The algebraic structure of the tetraquark-meson and tetraquark-diquark vertices $\Phi_a$ depend on the respective quantum numbers of the investigated four-quark state. For scalar four-quark states and (pseudo)scalar ingredients, e.g., those amplitudes are flavour and colour singlets and Lorentz scalars, otherwise they are Lorentz vectors or tensors.

\begin{figure}
	\includegraphics[width=8.8cm]{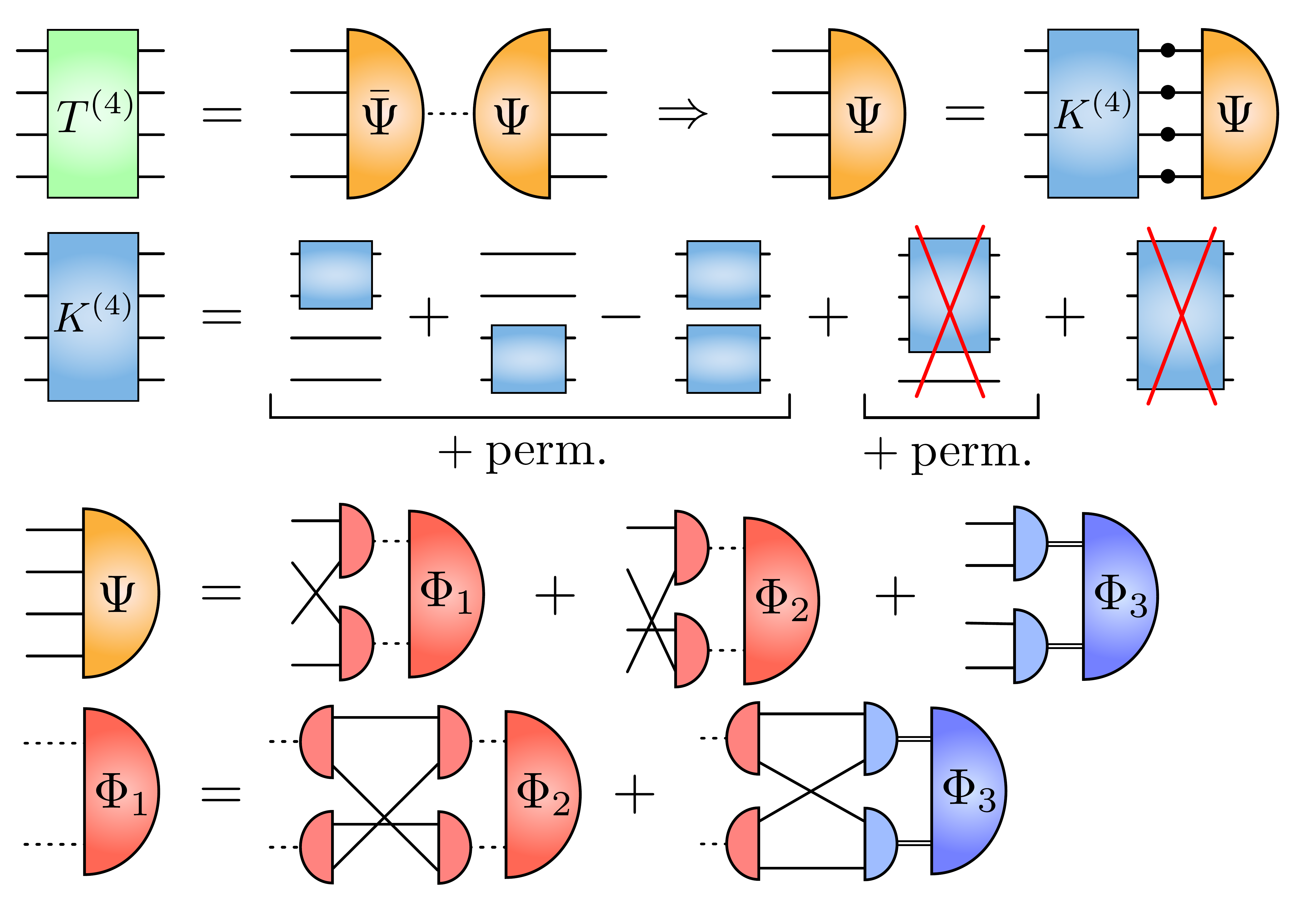}
	\caption{Diagrammatic representation of the basic quantities used in deriving the pure two-body/four-quark BSE.
		In the first line we display the representation of the bound state together with its corresponding BSE (\ref{bse}). In the second line we give the explicit decomposition of the interaction kernel $K^{(4)}$ in terms of irreducible two-, three- and four-body interactions. The red crosses indicate truncations, justified and explained in the main text. The third line displays the reduction of the four-body amplitude into a sum of two-body amplitudes featuring internal mesons (dashed lines) and \mbox{(anti-)}diquarks (double lines). One of the resulting effective two-body equations is given in the lowest line. The other two equations are obtained under permutations in the index set $\{1,2,3\}$, thus spanning the whole coupled system of two-body BSEs (for equal quark masses $\Phi_1$ and $\Phi_2$ are identical).\label{fig-diagram1}}
\end{figure}

With Eq.~(\ref{eq-4bsastructure}), we effectively solve for the vertices $\Phi_a$ while making use of solutions of the two-quark BSE for the amplitudes $\Gamma_{ij}$. The interaction kernel elements for the internal vertices $\Phi_a$ are quark exchange diagrams as visualized in the last line of Fig. \ref{fig-diagram1}.

\section{\label{sec:3} Technical details}

\subsection{\label{sec:3.1} Quark propagator, mesons and diquarks}
In order to solve the coupled four-quark two-body BSE, we need knowledge about other Green's functions such as quark propagators, meson and diquark amplitudes and propagators. These quantities have to be precalculated by solving the corresponding equations of motion, namely the quark Dyson-Schwinger equation (DSE) and the meson and diquark BSEs.
\\[4pt]
\textbf{Quark propagators.~~}
The quark DSE follows from the 1PI effective action via functional derivatives with respect to quark fields and reads

\begin{equation}
	S_{\alpha\beta}^{-1}(p)=Z_2(\ii\pslash+Z_mm_0)_{\alpha\beta}+C_F\int_q\mathcal{K}_{\alpha\alpha'\beta'\beta}S_{\alpha'\beta'}(q),
	\label{eq-qdse}
\end{equation}

with the wave-function and mass renormalization constants $Z_2$ and $Z_m$, the bare quark mass $m_0$ and the 
Casimir $C_F=4/3$ for $N_C=3$. The interaction kernel $\mathcal{K}$ contains the dressed gluon propagator 
as well as one bare and one dressed quark-gluon vertex. The Greek super-indices refer to colour, flavour 
and Dirac structure. We apply the Rainbow-Ladder approximation which proved to be reliable for ground state 
properties in all channels used in this work, see the review \cite{Eichmann:2016yit} for a detailed discussion. 
In this approximation, the kernel can be written as
\begin{equation}
\mathcal{K}_{\alpha\alpha'\beta'\beta}=Z_2^2\:\frac{4\pi\alpha(k^2)}{k^2}\:T^{\mu\nu}_k\:\gamma_{\alpha\alpha'}^\mu\:\gamma_{\beta\beta'}^\nu,
\label{eq-RL}
\end{equation}
where $T^{\mu\nu}_k$ is the transverse projector with respect to the momentum $k$ and the effective coupling $\alpha(k^2)$ occurs as the quantity which carries the non-trivial, momentum-dependent part of the gluon propagator and the quark-gluon vertex. The applied model for $\alpha(k^2)$ was taken from~\cite{Maris:1999nt} and has been discussed in detail e.g. in \cite{Eichmann:2016yit}.
\\[4pt]
\textbf{Meson and diquark amplitudes.~~} For mesons and diquarks one needs to solve the two-quark Bethe-Salpeter equation, i.e. Eq.~(\ref{bse}) for $n=2$. For the interaction kernel $K^{(2)}$ we adopt the same form as in the quark DSE, Eq.~(\ref{eq-RL}),
\begin{equation}
	K^{(2)} = \mathcal{K},
\end{equation}
as it preserves chiral symmetry through the axial-vector Ward-Takahashi identity~\cite{Eichmann:2016yit}. The 
amplitudes $\Gamma(P,p)$ occur as solutions of the BSEs and can be written as a linear combination of Dirac 
basis elements $\tau^{(\mu)}_i$ as follows (colour and flavour structure suppressed and $\mu$ occurs as a 
Lorentz index for $J=1$ states):
\begin{equation}
	\Gamma^{(\mu)}(P,p) = \sum_i \tau_i^{(\mu)}(P,p) f_i(P,p)
\end{equation}
For pseudoscalar mesons and scalar diquarks, there are four linear independent tensor structures and for vector 
mesons and axialvector diquarks there are eight. We solve the two-quark meson and diquark BSEs for the full 
set of basis elements and will approximate the solution by only taking into account the leading part of the 
amplitude, i.e.
\begin{equation}
	\Gamma^{(\mu)}(P,p) \approx \tau_1^{(\mu)}(P,p) f_1(P,p),
	\label{eq-bsa-approximation}
\end{equation}
which corresponds e.g. to $\tau_1=\gamma^5$ part for pseudoscalar mesons and $\tau_1^\mu=\gamma^\mu_\textrm{T}$ 
part for vector mesons~\cite{Williams:2018adr}, where the index T stands for the transverse projection with 
respect to the total meson momentum. For scalar and axialvector diquarks, one has to multiply the charge 
conjugation matrix $\mathcal{C}=\gamma^0\gamma^2$ on the tensor structures of pseudoscalar and vector BSEs. 
Only taking into account the leading structures of the respective amplitudes has proven to be a suitable, 
qualitative approximation of the full amplitude~\cite{Santowsky:10.22029}.

Mesons and diquarks with different flavour content could be calculated by taking quark propagators with 
different input quark masses $m_0$ (cf.~(\ref{eq-qdse})) as ingredients of the meson BSE (\ref{bse}). 
The input quark masses $(u,d,c)$ are fixed by ensuring that (i) the pion mass matches the (averaged) experimental 
value, (ii) the kaon mass is accurate and (iii) the sum $m_D + m_{D^*}$ agrees with experiment. We then
arrive at 
\begin{align}
	m_{0,u/d}=3.8\MeV && m_{0,s}=85.5\MeV && m_{0,c}=795\MeV
\end{align}
\begin{align}
	m_\pi 	&= 0.138\GeV 	& m_K 		&= 0.499\GeV \notag\\
	m_D 	&= 1.805\GeV 	& m_{D^*} 	&= 2.070\GeV \label{eq-dd-charm-mass}
\end{align}
Note that there is a certain mismatch between the heavy-light states and the charmonia. 
In order to obtain the experimental value of the $J/\psi$ meson,
\begin{align}
	m_{J/\psi}=3.10\GeV,
	\label{eq-jp-charm-mass}
\end{align}
a charm quark mass of 845\MeV{} is needed. This mismatch of 5\% reflects a systematic model error. 
Furthermore note that the heavy-light meson masses and amplitudes (in particular those of the $D$ mesons) are extrapolated due to quark poles in the integration domain~\cite{Windisch:2016iud}. 
\\[4pt]
\textbf{Meson and diquark propagators.~~}
For an exact description of the meson and diquark propagators, we generalize the $T$-matrix pole ansatz~(\ref{t-matrix-ansatz}) and replace the pole by a (potentially off-shell) propagator $D$:
\begin{equation}
	T^{(2n)}\approx\Gamma(P) D(P^2) \bar\Gamma(P)
\end{equation}
The propagator could then be calculated straightforwardly by using the solutions of the corresponding BSE via (\ref{t-matrix})~\cite{Eichmann:2016yit}.
\subsection{\label{sec:3.2} Solving a BSE in the presence of decay thresholds}\label{complex}
We solve a BSE by attaching an artificial eigenvalue function $\lambda(P^2)$ on the left hand side of Eq.~(\ref{bse}):
\begin{equation}
	\lambda(P^2)\:\Psi^{(n)} = K^{(n)} G_0^{(n)} \Psi^{(n)}
	\label{eq-bse-with-evcurve}
\end{equation}
The BSE is solved for a $P^2$ where $\lambda(P^2)=1$ holds. For bound states, the mass could then be extracted via $M^2=-P^2$. 
In the coupled system of BSEs as displayed in Fig.~\ref{fig-diagram1} the accessible region for the eigenvalue curve 
$\lambda(P^2)$ for real momenta $P^2$ is in principle restricted by the appearance of the first cut, i.e. the physical decay 
threshold. As an example, consider calculating a candidate for the $\chi_{c1}(3872)$ which may consist of $J/\psi$ and $\omega$ 
internally with their total momenta $P_{J/\Psi}$ and $P_\omega$.  With real relative momentum $p$ between these two mesons, these
momenta can be parametrized by $P_{J/\Psi} = p + \eta P$ and $P_{\omega} = -p + (1-\eta) P$ such that $P = P_{J/\Psi} + P_\omega$ 
for arbitrary $p$ and momentum is conserved. The partitioning parameter $\eta$ can then be adjusted such that the meson propagators 
are not probed on-shell until the physical decay threshold $M_\omega + M_{J/\Psi}$ is reached. In practise, however we face 
a somewhat tighter constraint for technical reasons: In order to be able to carry out the calculation, we have to
provide the (off-shell) Bethe-Salpeter amplitudes of the $J/\psi$, the $\omega$ and the heavy-light $D$-mesons after
quark exchange. The corresponding Bethe-Salpeter equations for the heavy-light mesons can be solved routinely for real 
relative momenta between the constituent quarks and antiquarks, but are currently out of reach for complex relative momenta. 
It is straight forward to show that this technical constraint leads to $\eta=1/2$ in the distribution of total momentum $P$
onto the total momenta of the $P_{J/\Psi}$ and $P_\omega$. This in turn leads to the constraint $P^2 > -4M_\omega ^2$ such that
the pole of the propagator of the omega-meson is avoided. In general, the constraint is $P^2 > -4m ^2$, where $m$ is the mass
of the lightest meson in the system of equations. For this region, we calculate the eigenvalue curve and 
extrapolate from there into the non-accessible region in order to obtain an on-shell solution for the four-quark state.
A similar procedure has been applied in Ref.~\cite{Santowsky:2020pwd} in the light quark sector. 

In principle, it would be very interesting to extract not only the mass of the bound state/resonance, but also
its Bethe-Salpeter wave function. The relative normalised weight of the different components of this wave function
would then allow for a direct and quantitative determination of the size of different contributions, i.e. 
meson-meson, hadro-charmonium or diquark-antidiquark. Unfortunately, such an extrapolation is much more complicated
as the above discussed extrapolation of the eigenvalue, since it would have to be done for every relative
momentum and consequently would face much larger uncertainties. We will study this possibility in future work.

\section{\label{sec:4} Results}
In this work we present results for the following flavour decompositions,
\begin{itemize}
	\item charmonium-like candidates with hidden charm ($c\bar{c}q\bar{q}$)
	\item open-charm states ($cc\bar{q}\bar{q}$)
	\item all-charm states ($cc\bar{c}\bar{c}$).
\end{itemize}
According to the two-body approximation, Eq.~(\ref{eq-4bsastructure}), 
we take into account different internal clusters. If the state in question has an experimental candidate, the choice
of our mesonic internal clusters are motivated by leading decay channels; if there is no such candidate, we choose the 
channels with lowest mass and with vanishing orbital momentum (as the two-body ansatz assumes $L=0$). For the
diquark-antidiquark clusters we also chose the combination with the lowest possible masses allowed by quantum 
numbers. Overall, we probe states with scalar and axialvector quantum numbers,
$J^P=0^+$ and $1^+$, with internal clusters that carry pseudo-scalar and vector quantum numbers for mesons
and scalar (S) and axialvector (A) quantum numbers for diquarks. These are precisely the channels where the underlying
rainbow-ladder approximation of the quark-gluon-interaction is known to work well \cite{Williams:2015cvx}. 
Other quantum numbers of the four-quark state require internal meson and diquark channels that are not well 
represented by this truncation and therefore no high quality results can be expected. We therefore postpone 
the study of such states to future work. 

We also like to point out that the current framework is only able to investigate the effects of internal 
two-body clusters that do not change the overall quark-content of the state in questions. This excludes the
formation of internal structures such as three-body $DD\pi$-components, which in some cases may have a 
considerable impact. This has been studied, e.g. for the $\chi_{c1}(3872)$ \cite{Baru:2011rs} and the 
$T_cc^+$ \cite{LHCb:2021auc,Du:2021zzh}. Including these effects would require a substantial generalisation 
of our framework which seems out of reach for now.

\subsection{\label{sec:4.1} Charmonium-like four-quark candidates with quark content $\boldsymbol{c\bar{c}q\bar{q}}$}
\begin{figure*}[t!]
	\includegraphics[width=8.8cm]{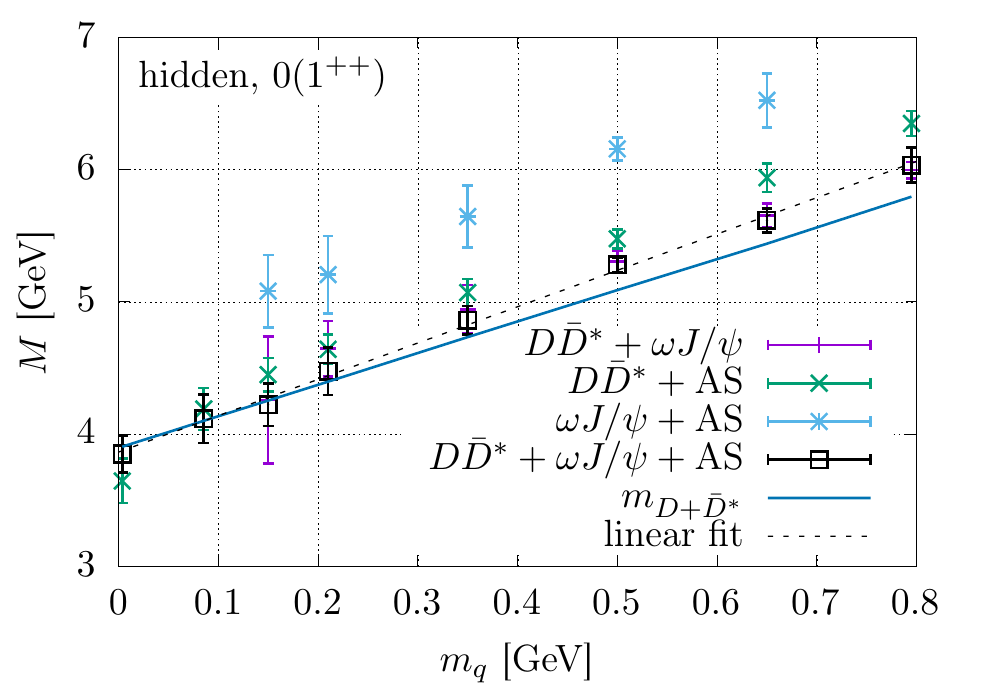}
	\includegraphics[width=8.8cm]{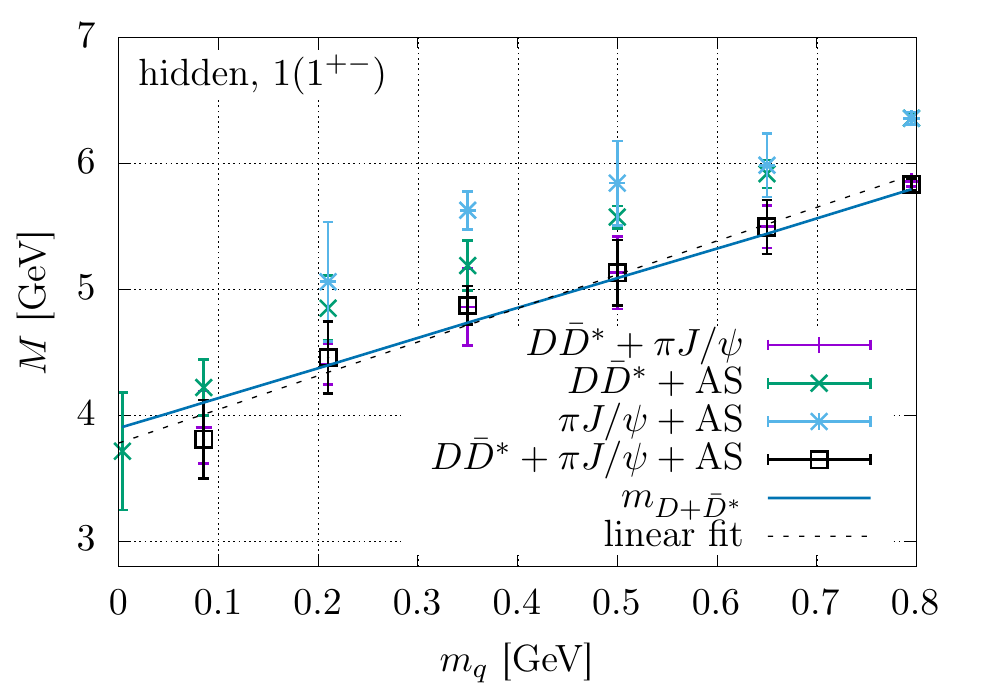}
	\includegraphics[width=8.8cm]{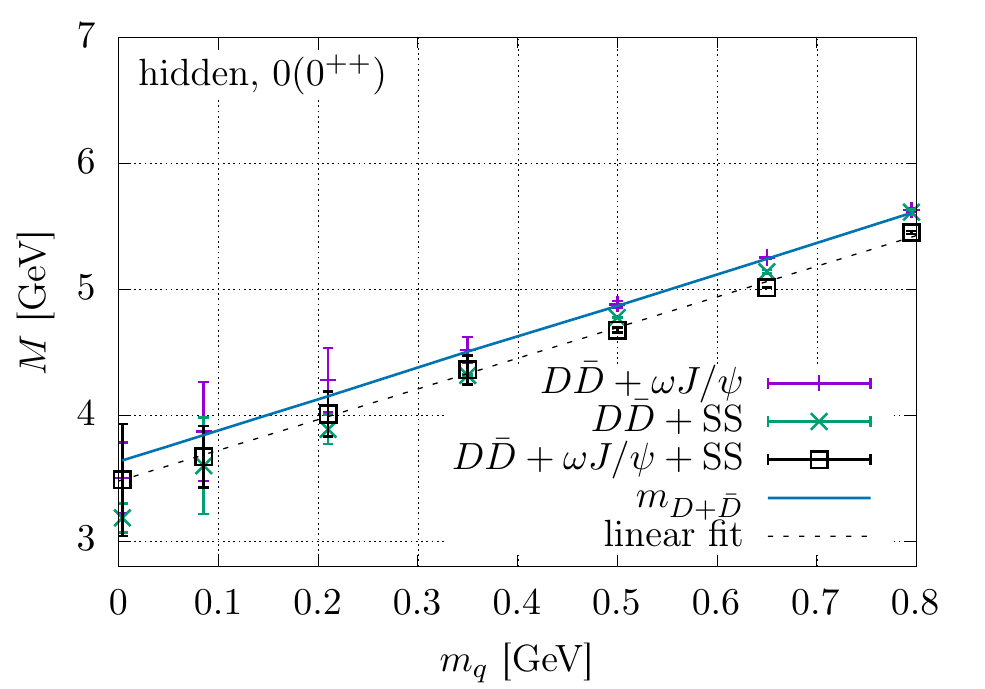}
    \includegraphics[width=8.8cm]{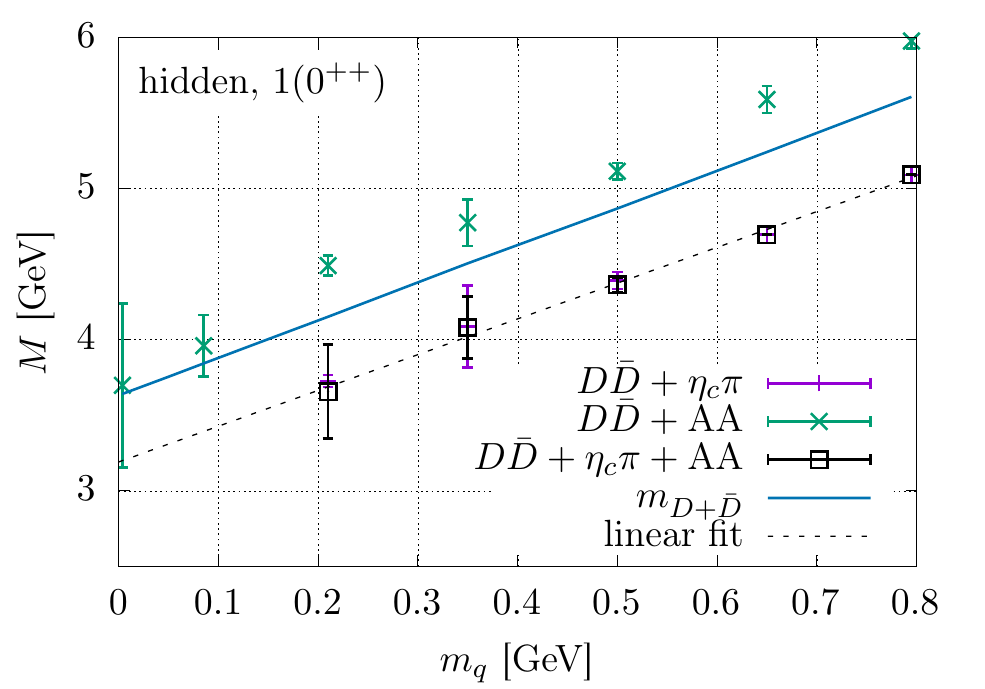}
	\caption{The mass curves for charmonium-like hidden-charm four-quark candidates with quantum numbers $0(1^{++})$ (upper left panel)
		$1(1^{+-})$ (upper right panel),  $0(0^{++})$ (lower left panel) and 
		$1(0^{+-})$ (lower right panel)with quark content $c\bar{c}q\bar{q}$ and dynamic light quark mass $m_q$.
		$A$ and $S$ stand for axialvector and scalar (anti-)diquarks. The short, blue bars on the left hand side denote the $D\bar{D}^{(*)}$ threshold at the physical point, $m_q=3.8\MeV$. }
	\label{fig-hidden}
\end{figure*}
We investigated charmonium-like tetraquarks for different quantum numbers in order to probe the experimentally 
confirmed axialvector states, $\chi_{c1}(3872)$ and $Z_c(3900)$, and further scalar states which are not yet 
confirmed. We show the ground state masses in Tab.~\ref{tab-hidden}. Variations in mass with the internal light 
quark masses are shown in Fig.~\ref{fig-hidden}. Here we also compare the full calculations, including all channels 
in question, with calculations including only part of the channels in order to identify the most dominant ones.

\begin{table}[t]
	\begin{tabular}{cccc}
		\hline
		$I(J^{PC})$ & \multicolumn{1}{l}{exp. candidate} & clusters                                  & \multicolumn{1}{l}{mass {[}GeV{]}} \\ \hline \hline
		$0(0^{++})$ & --                                 & $\boldsymbol{D\bar D} + \omega\:J/\psi + SS$   & 3.49(25)                                        \\
		$1(0^{++})$ & --                                 & $\boldsymbol{D\bar D} + \pi\:\eta_c + SS$      & 3.20(31)                                        \\
		$0(1^{++})$ & $\chi_{c1}(3872)$                  & $\boldsymbol{D\bar D^*} + \omega\:J/\psi + AS$ & 3.85(18)                                        \\
		$1(1^{+-})$ & $Z_c(3900)$                        & $\boldsymbol{D\bar D^*} + \pi\:J/\psi + AS$    & 3.79(31)                                        \\ \hline
	\end{tabular}
	\caption{Ground state masses for hidden-charm four-quark states with a pair of light quarks. The errors stem from extrapolations of the eigenvalue curve on the real axis. Bold-written clusters denote the dominant component in the equation. $A$ and $S$ stand for axialvector and scalar (anti-)diquarks.}
	\label{tab-hidden}
\end{table}
\begin{figure*}[t!]
	\includegraphics[width=8.8cm]{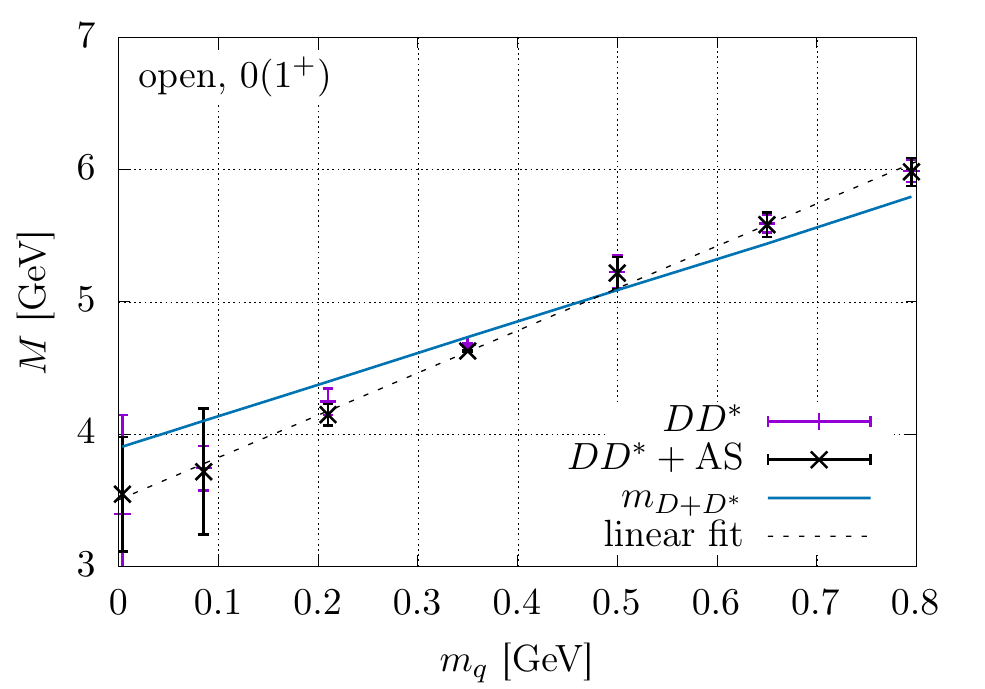}
	\includegraphics[width=8.8cm]{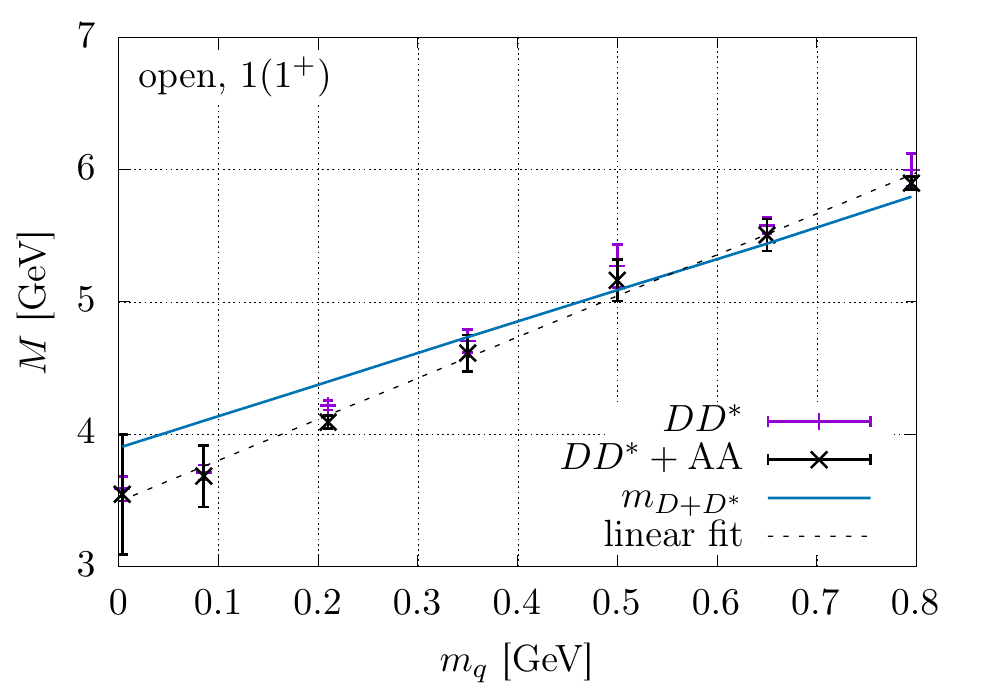}
	\includegraphics[width=8.8cm]{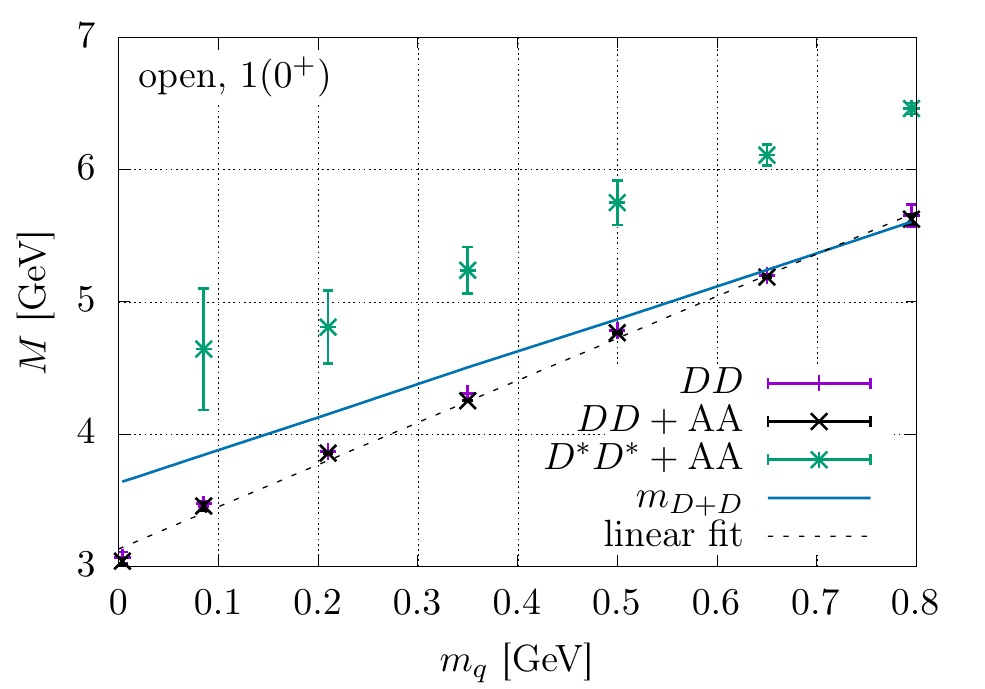}
	\caption{The mass curves for open-charm four-quark candidates and quantum numbers $0(1^{+})$ (upper left panel),
		$1(1^{+})$ (upper right panel) and $1(0^{+})$ (lower panel) with quark content $cc\bar{q}\bar{q}$ 
		and variations in the light quark mass $m_q$. The short, blue bars on the left hand side denote the $DD^{(*)}$ threshold at the physical point, $m_q=3.8\MeV$.}
	\label{fig-open}
\end{figure*}

For both quantum numbers, $0(1^{++})$ (left panel) and $1(1^{+-})$ (right panel), we observe that leaving out
$DD^*$ clusters changes the results dramatically, whereas they are hardly affected by the removal of the diquark 
clusters. We therefore conclude that the heavy-light meson-meson component is dominating and the diquark components
are negligible. The hadro-charmonium component has only a small impact on the $0(1^{++})$-state, whereas its 
contribution on the $1(1^{+-})$-state is much more relevant. This can be traced back to the presence
of a pion in the hadro-charmonium component of the isovector $1(1^{+-})$-state, which makes this component 
lighter and therefore more relevant than the corresponding hadro-charmonium component with $\omega$ in the 
isoscalar $0(1^{++})$-state.
Overall, the masses of both experimental candidates, the $\chi_{c1}(3872)$ and the $Z_c(3900)$, are reproduced 
successfully by our calculations within error bars. Whether these states are below or above threshold could 
not be resolved within the errors of our calculations. 
Note that we are therefore not in a position to prove 
(or disprove) a potential molecular nature of these states: while the dominance of heavy-light meson components 
in their wave functions are certainly compatible with (and even may suggest) a molecular nature, we cannot pin 
down the required small binding energy with sufficient accuracy.

On the other hand, we find four-quark states with scalar angular momentum well below the threshold. Thus they 
could very well be bound states. They are also dominated by the heavy-light meson-meson components consisting 
of $D$ and $\bar{D}$ mesons, since without the $D\bar{D}$-clusters we do not obtain any solutions of the BSEs.
Again we observe that the diquarks are negligible in both cases and the lighter hadro-charmonium component
in the isovector state is more important than the heavy hadro-charmonium component in the isoscalar one. 

The ground state masses in Tab.~\ref{tab-hidden} have been extracted by fitting a linear curve (dashed) to the
mass curves. Within error bars, this linear fit seems to work very well and provides a rough estimate of the
(real part of the) masses in our approach. 
In general, the mass ordering between the scalars and the axialvectors is natural, as axialvector states 
are more massive than their scalar counterparts, and agrees with the one expected in a molecule picture
\cite{Cleven:2015era,Guo:2017jvc}. Note that in contrast to lattice QCD, calculations at the physical 
point in our approach do not require more resources than for heavier quark masses. For the isovector 
candidates however, there are additional technical complications due to the occurrence of pion poles
at small time-like momenta. Since in the present set-up we cannot go beyond these poles, the necessary 
extrapolations to the pole location of the four-quark state have to bridge an enormous mass range resulting
in very large error bars and cease to be useful. In the plots we only show results at the physical point 
with an excluded hadro-charmonium component, because in that case the problematic pion pole is absent.

Running up the mass curves for isoscalar states from $m_{u/d}\rightarrow m_s$, it is possible to extract also candidates including strange instead of light quarks. The corresponding masses are given by
\begin{align}
	M_{c\bar{c}s\bar{s}, 0^{++}} &= 3.69(18)\GeV\notag \\ M_{c\bar{c}s\bar{s}, 1^{++}} &= 4.10(16)\GeV,
\end{align}
which makes it possible to identify the axialvector candidate with the $\chi_{c1}(4140)$ which decays at least into $\Jp\:\phi$, although the dominant internal structure in our calculations is $D_s\bar D_s^*$.

Our results agree very well with recent four-body calculations on a quantitative
level~\cite{Wallbott:2019dng,Wallbott:2020jzh}. In particular our main findings, heavy-light meson dominance
and diquark suppression are similar. Thus, both approaches are consistent with each other.

\subsection{\label{sec:4.2} Open-charm states with quark content $\boldsymbol{cc\bar{q}\bar{q}}$}
Similar to section~\ref{sec:4.1} we also investigated open-charm states with quark content $cc\bar{q}\bar{q}$. 
The corresponding mass curves are shown in Fig.~\ref{fig-open} and the extrapolated masses in Tab.~\ref{tab-open}. 
Note that the open charm states here underlie 
Pauli symmetry, which restricts the choice of diquark-antidiquark components as they stem from the colour-antitriplet.
In particular, this is the reason why there is no isoscalar-scalar state in our approximation and why the lowest-lying
allowed diquark-antidiquark cluster consists only of axialvector diquarks, $AA$, in the scalar case. 

\begin{table}[t]
	\begin{tabular}{cccc}
		\hline
		$I(J^{P})$ & \multicolumn{1}{l}{exp. candidate} & clusters                          & \multicolumn{1}{l}{mass {[}GeV{]}} \\ \hline
		$1(0^{+})$ & --                        			& $\boldsymbol{DD} + D^*D^* + AA$   & 3.21(2)                                         \\
		$0(1^{+})$ & $T^+_{cc}$                         & $\boldsymbol{DD^*} + D^*D^* + AS$ & 3.49(48)                                        \\
		$1(1^{+})$ & --                                 & $\boldsymbol{DD^*} + D^*D^* + AA$ & 3.47(24)                                        \\ \hline
	\end{tabular}
	\caption{Ground state masses for open-charm four-quark states. The errors stem from extrapolations of the eigenvalue curve on the real axis. Bold-written clusters denote the dominant component in the equation. $A$ and $S$ stand for axialvector and scalar (anti-)diquarks.}
	\label{tab-open}
\end{table}

We observe that both axialvector four-quark states are (slightly) higher in mass than the scalar state, 
which is consistent with our findings in the hidden-charm sector. Different from the hidden-charm case, 
the masses of the isoscalar and isovector states are in a similar mass region, which may be traced back 
to the absence of an influential hadro-charmonium component. Unfortunately, and similar to corresponding 
lattice calculations \cite{Ikeda:2013vwa,Cheung:2017tnt,Junnarkar:2018twb}, we cannot say for certain 
that our axialvector states are bound. This seems to be different for the scalar state, where the extrapolation 
errors are small enough to suggest a bound state at the physical point. 

Currently, there is only one experimental candidate for an open-charm state, the recently discovered isoscalar 
axialvector $T_{cc}^+$ with a mass of $\sim3875\MeV$ and a leading decay channel compatible with an inner 
composition of $DD^*$~\cite{LHCb:2021vvq,LHCb:2021auc}. Within error bars, the experimental mass and our 
mass agrees, even as our mean value is about ten percent too small. We come back to this discussion 
in section~\ref{sec:5}. The almost mass degenerate axialvector isovector state has been searched for,
but not found by the LHCb collaboration in the respective channels \cite{LHCb:2021auc}. This is an interesting
observation that deserves further consideration. 

The axialvector masses for the open-charm states in our framework are smaller than the hidden-charm equivalents. 
An explanation for this is the corresponding interaction on the level of the two-body BSE. Due to the absence of 
a hadro-charmonium component the dominating interaction diagram is the quark exchange between two identical 
$D$ meson clusters, cf.~Fig.~\ref{fig-diagram1}. As it turns out, the $D$ meson Bethe-Salpeter amplitudes 
are larger than the ones of \Jp{} and $\omega$ or \Jp{} and $\pi$. This implies a higher eigenvalue
curve~(\ref{eq-bse-with-evcurve}) and therefore, a lighter ground state. This is in agreement with
early studies of open flavour heavy-light systems \cite{Manohar:1992nd}. 

%
%
%
\subsection{\label{sec:4.3} All-charm states with quark content $\boldsymbol{cc\bar{c}\bar{c}}$}

\begin{figure}[t]
	\centering
	\includegraphics[width=0.48\textwidth]{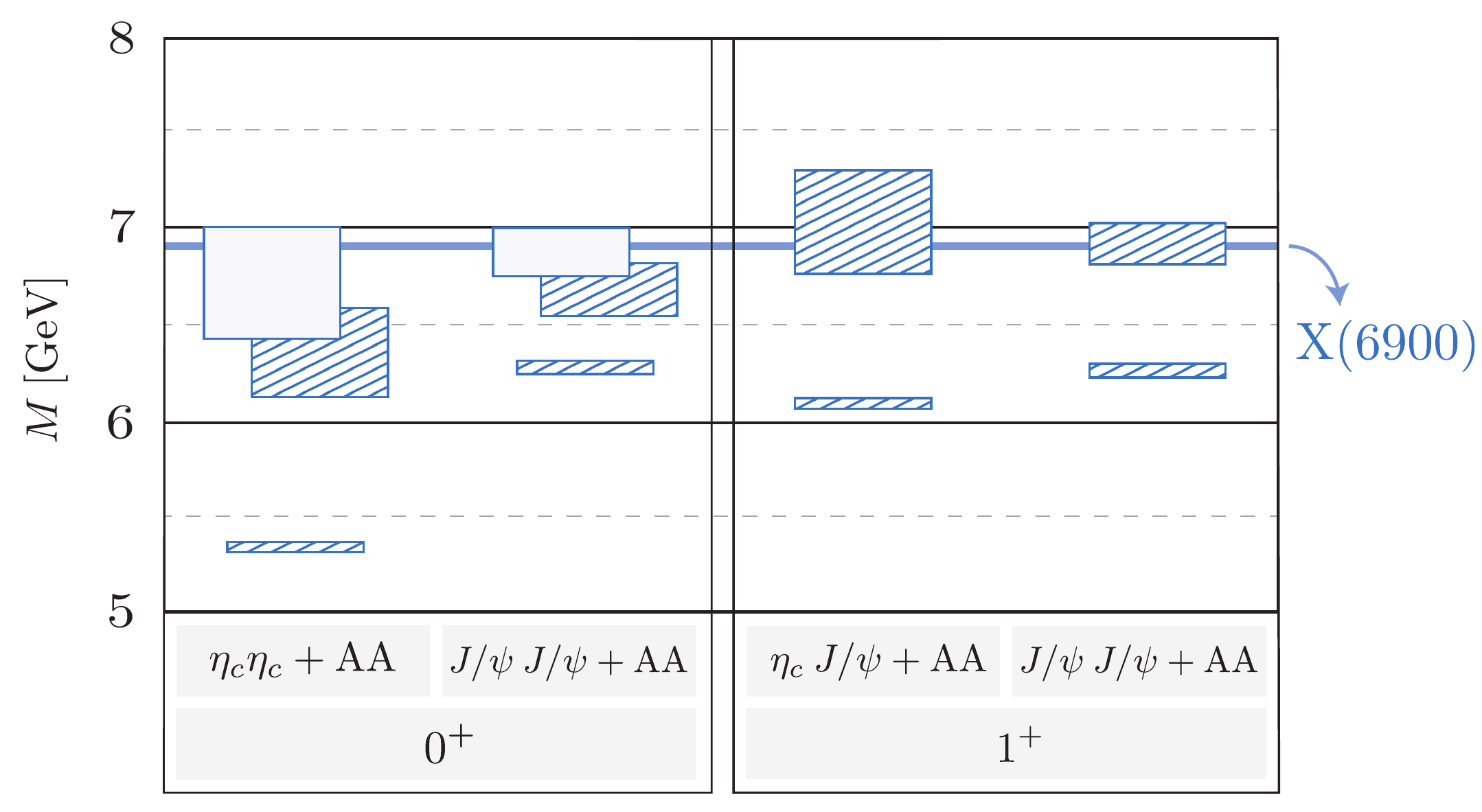}
	\caption{The all-charm spectrum showed graphically with different quantum numbers $0^+$ and $1^+$, showing the ground states and excited ones. 
		Striped rectangles denote states which are dominated strongly by the meson-meson components, such that diquark-antidiquark components are 
		completely negligible. In contrast, open rectangles denote states with still dominant meson-meson components, but significant 
		diquark-antidiquark admixtures. The vertical extent of the rectangles denote the numerical error from the extrapolation of the 
		corresponding eigenvalue curve and the continuous, horizontal blue line denotes the experimental value for the X(6900)~\cite{LHCb:2020bwg}.}
	\label{fig-allcharm-spec}
\end{figure}
The quantum numbers of the only experimentally confirmed candidate for an all-charm state, the $X(6900)$, have not yet
been identified. It is therefore reasonable to probe different quantum numbers with different internal configurations 
in order to search for potential candidates. Similar to the open-charm case, Pauli symmetry restricts the quantum numbers 
of the (anti-)diquarks, i.e. explains why only axialvector diquarks could occur as possible clusters. Technically,
all-charm states offer a unique possibility within the restrictions of our current framework: in the accessible 
region of total momenta not only the eigenvalue curves for the ground states but also those for the radially excites states
are not too far away from the physical point, $\lambda=1$, it is possible to use our well-probed extrapolation procedure 
to study the excitation spectra of these candidates as well. The different states we consider are shown in Tab.~\ref{tab-allcharm} 
along with the calculated masses. Fig.~\ref{fig-allcharm-spec} shows the calculated spectrum graphically.

\begin{table}[t]
	\begin{tabular}{ccccc}
		\hline
		$J^{P}$                  & clusters              & $M$ {[}GeV{]} & $M^*$ {[}GeV{]} & $M^{**}$ {[}GeV{]} \\ \hline\hline
		\multirow{2}{*}{$0^{+}$} & $\boldsymbol{\eta_c\eta_c} + AA$   & 5.34(2)       & 6.30(13)        & 6.70(30)           \\
		& $\boldsymbol{J/\psi\:J/\psi} + AA$ & 6.30(3)       & 6.71(14)        & 6.87(12)           \\ \hline
		\multirow{2}{*}{$1^{+}$} & $\boldsymbol{\eta_c\:J/\psi} + AA$ & 6.07(2)       & 7.03(26)        & --                 \\
		& $\boldsymbol{J/\psi\:J/\psi} + AA$ & 6.28(4)       & 6.92(12)        & --                 \\ \hline
	\end{tabular}
	\caption{Numerical values for the ground states and the excitation spectra (excitation levels denoted by stars, $*$) of different all-charm four-quark candidates using different clusters. Bold-written clusters are the dominant ones. $A$ stands for an axialvector (anti-)diquark.}
	\label{tab-allcharm}
\end{table}

We find that all ground states are too low in mass to provide a proper description of the $X(6900)$. Instead, we find possible states 
within the excitation spectra for both quantum numbers, $0^+$ and $1^+$. Whereas potential candidates are found as second radial excitations 
of the scalar states, the axialvector ones occur as first excitations. Only the ground state of the di-\nc{} state is surely bound whereas 
all other states are either clearly above or in the proximity of the corresponding, lowest-lying decay threshold. Without exceptions, the 
states are dominated by the meson-meson component; only the second excitations (**) in the scalar spectrum have significant (although still small) 
diquark-antidiquark components. Note that, again, meson-meson dominance alone is not sufficient to conclude anything 
about the potential molecular nature of these states (cf. the discussion of the $\chi_{c1}(3872)$ and the 
$Z_c(3900)$ above). 

In the $1^+$ case, both configurations, the first excitations of \nc\:\Jp{} and \Jp\:\Jp, overlap with the 
experimental state within error bars. Thus, it is possible that the experimental state is a mixture of these configurations. In our calculations, we have not yet studied this possibility due to the associated complexity. 
This is left for future studies. 

Many previous model calculations of the $X(6900)$ assume a diquark-antidiquark structure 
\cite{Giron:2020wpx,Faustov:2020qfm,Faustov:2021hjs,Zhao:2020nwy,Deng:2020iqw} and find a first radial excitation in the mass region of the experimental 
state. While we agree with theses studies on the general notion that the experimental state is a radial excitation, our findings seem to invalidate
diquark-models for these states on general grounds. Our findings partly agree, however, with very recent calculations in a non-relativistic quark model that uses 
a spin-independent Cornell potential based on lattice calculations to investigate meson-meson resonances \cite{Yang:2021hrb}. The agreement is especially 
present in the $1^+$ channel, where the $J/\psi\:J/\psi$ ground and excited state are in a very similar mass region. This further supports the interpretation 
of the experimental state as an axialvector di-$J/\psi$ resonance.
\section{\label{sec:5} Concluding Remarks}\label{sec:sum}
In this work we studied the inner structure of a number of four-quark candidates with charm quarks. These involve charmonium-like hidden-charm states 
with quark content $c\bar{c}q\bar{q}$, open-charm states with quark content $cc\bar{q}\bar{q}$ and all-charm states with quark content $cc\bar{c}\bar{c}$. 
In a parameter free calculation we probed different quantum numbers and found reasonable descriptions of experimentally confirmed states, 
dominated by internal meson-meson configurations. In the sector of charmonium-like states, the $\chi_{c1}(3872)$ and the $Z_c(3900)$ are 
well-described as $D\bar D^*$ states. Furthermore we find a $D_s\bar D_s^*$ dominated state which may be identified with the $\chi_{c1}(4140)$. 
In the open-charm region we found an isoscalar axialvector $DD^*$ state which may be identified with the recently discovered $T_{cc}^+$. In the all-charm 
sector, we find possible candidates for the $X(6900)$ in the excitation spectra of scalar and axialvector four-quark states. We were also able 
to make predictions for the inner structure of many additional states which have not yet been experimentally confirmed. In general, internal 
diquark-antidiquark configurations are always found to be subleading and in many cases even negligible. This invalidates models based on diquark
degrees of freedom on general grounds. Whenever possible due to quantum numbers we also see hadro-charmonium components, but again only as 
sub-dominant components of the full wave function. A possible mixing with ordinary charmonia which would be interesting for charmonium-like 
states with $I=0$ was not yet included for technical reasons.

Generally speaking, our results in this work are qualitative, mainly because the two-body equation is truncated (a) by neglecting three- and 
four-body forces, cf. Eq.~(\ref{eq-2-3-4-body-forces}), and (b) by only taking into account the leading component of the meson and diquark BSAs, 
cf. Eq.~(\ref{eq-bsa-approximation}). The systematic quantitative error following from those truncations is hard to estimate. From our experience
with other quantities, e.g. decay constants, where all components can be taken into account we infer that the potential error may be on the twenty 
percent level. Further uncertainties come from extrapolations of the eigenvalue curves as introduced in~(\ref{eq-bse-with-evcurve}). This error
has been quantified. All these sources of error result in inaccurately resolved masses, but we do not expect that  physically relevant qualitative 
aspects such as mass orderings from which we deduce dominant clusters are affected. In particular we are reasonably certain, that all sources of 
error due not affect the general statement that diquarks are mostly irrelevant. 

This work could also be considered as groundwork for further calculations. For a more complete understanding of the states in question it would 
be reasonable (a) to couple the four-quark components to a quark-antiquark state if the quantum numbers allow \cite{Santowsky:2020pwd} 
and (b) to further investigate the eigenvalue curve for complex $P^2$ in order to describe states as four-quark resonances including their widths, 
as done for the light quark sector in Ref.~\cite{Santowsky:2021ugd}. This is subject of future work and connected to solving further technical 
challenges such as gaining knowledge of $D$ meson amplitudes in the complex plane and the possibility to circumvent quark, meson and diquark 
poles dynamically while integrating. 

In any case, to our mind the results of this work demonstrate that functional continuum methods based on the Bethe-Salpeter equations of QCD 
are an adequate and systematic tool to decode the inner structure of exotic states with heavy quarks involved.
\subsection*{Acknowledgements}
We are grateful to Gernot Eichmann, Paul C. Wallbott and Marc Wagner for useful discussions. 
This work was supported by the Helmholtz Research Academy Hesse for FAIR (HFHF), 
by the GSI Helmholtzzentrum f\"{u}r Schwerionenforschung and by BMBF under grant number 05P21RGFP3.

\bibliographystyle{apsrev4-1}
\bibliography{complex_lights}

\begin{thebibliography}{56}%
\makeatletter
\providecommand \@ifxundefined [1]{%
 \@ifx{#1\undefined}
}%
\providecommand \@ifnum [1]{%
 \ifnum #1\expandafter \@firstoftwo
 \else \expandafter \@secondoftwo
 \fi
}%
\providecommand \@ifx [1]{%
 \ifx #1\expandafter \@firstoftwo
 \else \expandafter \@secondoftwo
 \fi
}%
\providecommand \natexlab [1]{#1}%
\providecommand \enquote  [1]{``#1''}%
\providecommand \bibnamefont  [1]{#1}%
\providecommand \bibfnamefont [1]{#1}%
\providecommand \citenamefont [1]{#1}%
\providecommand \href@noop [0]{\@secondoftwo}%
\providecommand \href [0]{\begingroup \@sanitize@url \@href}%
\providecommand \@href[1]{\@@startlink{#1}\@@href}%
\providecommand \@@href[1]{\endgroup#1\@@endlink}%
\providecommand \@sanitize@url [0]{\catcode `\\12\catcode `\$12\catcode
  `\&12\catcode `\#12\catcode `\^12\catcode `\_12\catcode `\%12\relax}%
\providecommand \@@startlink[1]{}%
\providecommand \@@endlink[0]{}%
\providecommand \url  [0]{\begingroup\@sanitize@url \@url }%
\providecommand \@url [1]{\endgroup\@href {#1}{\urlprefix }}%
\providecommand \urlprefix  [0]{URL }%
\providecommand \Eprint [0]{\href }%
\providecommand \doibase [0]{http://dx.doi.org/}%
\providecommand \selectlanguage [0]{\@gobble}%
\providecommand \bibinfo  [0]{\@secondoftwo}%
\providecommand \bibfield  [0]{\@secondoftwo}%
\providecommand \translation [1]{[#1]}%
\providecommand \BibitemOpen [0]{}%
\providecommand \bibitemStop [0]{}%
\providecommand \bibitemNoStop [0]{.\EOS\space}%
\providecommand \EOS [0]{\spacefactor3000\relax}%
\providecommand \BibitemShut  [1]{\csname bibitem#1\endcsname}%
\let\auto@bib@innerbib\@empty
\bibitem [{\citenamefont {Choi}\ \emph {et~al.}(2003)\citenamefont {Choi} \emph
  {et~al.}}]{Belle:2003nnu}%
  \BibitemOpen
  \bibfield  {author} {\bibinfo {author} {\bibfnamefont {S.~K.}\ \bibnamefont
  {Choi}} \emph {et~al.} (\bibinfo {collaboration} {Belle}),\ }\href {\doibase
  10.1103/PhysRevLett.91.262001} {\bibfield  {journal} {\bibinfo  {journal}
  {Phys. Rev. Lett.}\ }\textbf {\bibinfo {volume} {91}},\ \bibinfo {pages}
  {262001} (\bibinfo {year} {2003})},\ \Eprint
  {http://arxiv.org/abs/hep-ex/0309032} {arXiv:hep-ex/0309032} \BibitemShut
  {NoStop}%
\bibitem [{\citenamefont {Esposito}\ \emph {et~al.}(2016)\citenamefont
  {Esposito}, \citenamefont {Pilloni},\ and\ \citenamefont
  {Polosa}}]{Esposito:2016noz}%
  \BibitemOpen
  \bibfield  {author} {\bibinfo {author} {\bibfnamefont {A.}~\bibnamefont
  {Esposito}}, \bibinfo {author} {\bibfnamefont {A.}~\bibnamefont {Pilloni}}, \
  and\ \bibinfo {author} {\bibfnamefont {A.~D.}\ \bibnamefont {Polosa}},\
  }\href {\doibase 10.1016/j.physrep.2016.11.002} {\bibfield  {journal}
  {\bibinfo  {journal} {Phys. Rept.}\ }\textbf {\bibinfo {volume} {668}},\
  \bibinfo {pages} {1} (\bibinfo {year} {2016})},\ \Eprint
  {http://arxiv.org/abs/1611.07920} {arXiv:1611.07920 [hep-ph]} \BibitemShut
  {NoStop}%
\bibitem [{\citenamefont {Ali}\ \emph {et~al.}(2017)\citenamefont {Ali},
  \citenamefont {Lange},\ and\ \citenamefont {Stone}}]{Ali:2017jda}%
  \BibitemOpen
  \bibfield  {author} {\bibinfo {author} {\bibfnamefont {A.}~\bibnamefont
  {Ali}}, \bibinfo {author} {\bibfnamefont {J.~S.}\ \bibnamefont {Lange}}, \
  and\ \bibinfo {author} {\bibfnamefont {S.}~\bibnamefont {Stone}},\ }\href
  {\doibase 10.1016/j.ppnp.2017.08.003} {\bibfield  {journal} {\bibinfo
  {journal} {Prog. Part. Nucl. Phys.}\ }\textbf {\bibinfo {volume} {97}},\
  \bibinfo {pages} {123} (\bibinfo {year} {2017})},\ \Eprint
  {http://arxiv.org/abs/1706.00610} {arXiv:1706.00610 [hep-ph]} \BibitemShut
  {NoStop}%
\bibitem [{\citenamefont {Brambilla~et al.}(2020)}]{Brambilla:2019esw}%
  \BibitemOpen
  \bibfield  {author} {\bibinfo {author} {\bibfnamefont {N.}~\bibnamefont
  {Brambilla~et al.}},\ }\href@noop {} {\bibfield  {journal} {\bibinfo
  {journal} {Phys. Rept.}\ }\textbf {\bibinfo {volume} {873}},\ \bibinfo
  {pages} {1} (\bibinfo {year} {2020})},\ \Eprint
  {http://arxiv.org/abs/1907.07583} {arXiv:1907.07583 [hep-ex]} \BibitemShut
  {NoStop}%
\bibitem [{\citenamefont {Aaij}\ \emph
  {et~al.}(2021{\natexlab{a}})\citenamefont {Aaij} \emph
  {et~al.}}]{LHCb:2021vvq}%
  \BibitemOpen
  \bibfield  {author} {\bibinfo {author} {\bibfnamefont {R.}~\bibnamefont
  {Aaij}} \emph {et~al.} (\bibinfo {collaboration} {LHCb}),\ }\href@noop {} {\
  (\bibinfo {year} {2021}{\natexlab{a}})},\ \Eprint
  {http://arxiv.org/abs/2109.01038} {arXiv:2109.01038 [hep-ex]} \BibitemShut
  {NoStop}%
\bibitem [{\citenamefont {Aaij}\ \emph
  {et~al.}(2020{\natexlab{a}})\citenamefont {Aaij} \emph
  {et~al.}}]{LHCb:2020bwg}%
  \BibitemOpen
  \bibfield  {author} {\bibinfo {author} {\bibfnamefont {R.}~\bibnamefont
  {Aaij}} \emph {et~al.} (\bibinfo {collaboration} {LHCb}),\ }\href {\doibase
  10.1016/j.scib.2020.08.032} {\bibfield  {journal} {\bibinfo  {journal} {Sci.
  Bull.}\ }\textbf {\bibinfo {volume} {65}},\ \bibinfo {pages} {1983} (\bibinfo
  {year} {2020}{\natexlab{a}})},\ \Eprint {http://arxiv.org/abs/2006.16957}
  {arXiv:2006.16957 [hep-ex]} \BibitemShut {NoStop}%
\bibitem [{\citenamefont {Guo}\ \emph {et~al.}(2018)\citenamefont {Guo},
  \citenamefont {Hanhart}, \citenamefont {Mei\ss{}ner}, \citenamefont {Wang},
  \citenamefont {Zhao},\ and\ \citenamefont {Zou}}]{Guo:2017jvc}%
  \BibitemOpen
  \bibfield  {author} {\bibinfo {author} {\bibfnamefont {F.-K.}\ \bibnamefont
  {Guo}}, \bibinfo {author} {\bibfnamefont {C.}~\bibnamefont {Hanhart}},
  \bibinfo {author} {\bibfnamefont {U.-G.}\ \bibnamefont {Mei\ss{}ner}},
  \bibinfo {author} {\bibfnamefont {Q.}~\bibnamefont {Wang}}, \bibinfo {author}
  {\bibfnamefont {Q.}~\bibnamefont {Zhao}}, \ and\ \bibinfo {author}
  {\bibfnamefont {B.-S.}\ \bibnamefont {Zou}},\ }\href {\doibase
  10.1103/RevModPhys.90.015004} {\bibfield  {journal} {\bibinfo  {journal}
  {Rev. Mod. Phys.}\ }\textbf {\bibinfo {volume} {90}},\ \bibinfo {pages}
  {015004} (\bibinfo {year} {2018})},\ \Eprint
  {http://arxiv.org/abs/1705.00141} {arXiv:1705.00141 [hep-ph]} \BibitemShut
  {NoStop}%
\bibitem [{\citenamefont {Esposito}\ \emph {et~al.}(2022)\citenamefont
  {Esposito}, \citenamefont {Maiani}, \citenamefont {Pilloni}, \citenamefont
  {Polosa},\ and\ \citenamefont {Riquer}}]{Esposito:2021vhu}%
  \BibitemOpen
  \bibfield  {author} {\bibinfo {author} {\bibfnamefont {A.}~\bibnamefont
  {Esposito}}, \bibinfo {author} {\bibfnamefont {L.}~\bibnamefont {Maiani}},
  \bibinfo {author} {\bibfnamefont {A.}~\bibnamefont {Pilloni}}, \bibinfo
  {author} {\bibfnamefont {A.~D.}\ \bibnamefont {Polosa}}, \ and\ \bibinfo
  {author} {\bibfnamefont {V.}~\bibnamefont {Riquer}},\ }\href {\doibase
  10.1103/PhysRevD.105.L031503} {\bibfield  {journal} {\bibinfo  {journal}
  {Phys. Rev. D}\ }\textbf {\bibinfo {volume} {105}},\ \bibinfo {pages}
  {L031503} (\bibinfo {year} {2022})},\ \Eprint
  {http://arxiv.org/abs/2108.11413} {arXiv:2108.11413 [hep-ph]} \BibitemShut
  {NoStop}%
\bibitem [{\citenamefont {Aaij}\ \emph
  {et~al.}(2020{\natexlab{b}})\citenamefont {Aaij} \emph
  {et~al.}}]{LHCb:2020xds}%
  \BibitemOpen
  \bibfield  {author} {\bibinfo {author} {\bibfnamefont {R.}~\bibnamefont
  {Aaij}} \emph {et~al.} (\bibinfo {collaboration} {LHCb}),\ }\href {\doibase
  10.1103/PhysRevD.102.092005} {\bibfield  {journal} {\bibinfo  {journal}
  {Phys. Rev. D}\ }\textbf {\bibinfo {volume} {102}},\ \bibinfo {pages}
  {092005} (\bibinfo {year} {2020}{\natexlab{b}})},\ \Eprint
  {http://arxiv.org/abs/2005.13419} {arXiv:2005.13419 [hep-ex]} \BibitemShut
  {NoStop}%
\bibitem [{\citenamefont {Barucca}\ \emph {et~al.}(2019)\citenamefont {Barucca}
  \emph {et~al.}}]{PANDA:2018zjt}%
  \BibitemOpen
  \bibfield  {author} {\bibinfo {author} {\bibfnamefont {G.}~\bibnamefont
  {Barucca}} \emph {et~al.} (\bibinfo {collaboration} {PANDA}),\ }\href
  {\doibase 10.1140/epja/i2019-12718-2} {\bibfield  {journal} {\bibinfo
  {journal} {Eur. Phys. J.}\ }\textbf {\bibinfo {volume} {A55}},\ \bibinfo
  {pages} {42} (\bibinfo {year} {2019})},\ \Eprint
  {http://arxiv.org/abs/1812.05132} {arXiv:1812.05132 [hep-ex]} \BibitemShut
  {NoStop}%
\bibitem [{\citenamefont {Barucca}\ \emph {et~al.}(2021)\citenamefont {Barucca}
  \emph {et~al.}}]{PANDA:2021ozp}%
  \BibitemOpen
  \bibfield  {author} {\bibinfo {author} {\bibfnamefont {G.}~\bibnamefont
  {Barucca}} \emph {et~al.} (\bibinfo {collaboration} {PANDA}),\ }\href
  {\doibase 10.1140/epja/s10050-021-00475-y} {\bibfield  {journal} {\bibinfo
  {journal} {Eur. Phys. J. A}\ }\textbf {\bibinfo {volume} {57}},\ \bibinfo
  {pages} {184} (\bibinfo {year} {2021})},\ \Eprint
  {http://arxiv.org/abs/2101.11877} {arXiv:2101.11877 [hep-ex]} \BibitemShut
  {NoStop}%
\bibitem [{\citenamefont {Maiani}\ \emph {et~al.}(2005)\citenamefont {Maiani},
  \citenamefont {Piccinini}, \citenamefont {Polosa},\ and\ \citenamefont
  {Riquer}}]{Maiani:2004vq}%
  \BibitemOpen
  \bibfield  {author} {\bibinfo {author} {\bibfnamefont {L.}~\bibnamefont
  {Maiani}}, \bibinfo {author} {\bibfnamefont {F.}~\bibnamefont {Piccinini}},
  \bibinfo {author} {\bibfnamefont {A.~D.}\ \bibnamefont {Polosa}}, \ and\
  \bibinfo {author} {\bibfnamefont {V.}~\bibnamefont {Riquer}},\ }\href
  {\doibase 10.1103/PhysRevD.71.014028} {\bibfield  {journal} {\bibinfo
  {journal} {Phys. Rev.}\ }\textbf {\bibinfo {volume} {D71}},\ \bibinfo {pages}
  {014028} (\bibinfo {year} {2005})},\ \Eprint
  {http://arxiv.org/abs/hep-ph/0412098} {arXiv:hep-ph/0412098 [hep-ph]}
  \BibitemShut {NoStop}%
\bibitem [{\citenamefont {Ebert}\ \emph {et~al.}(2007)\citenamefont {Ebert},
  \citenamefont {Faustov}, \citenamefont {Galkin},\ and\ \citenamefont
  {Lucha}}]{Ebert:2007rn}%
  \BibitemOpen
  \bibfield  {author} {\bibinfo {author} {\bibfnamefont {D.}~\bibnamefont
  {Ebert}}, \bibinfo {author} {\bibfnamefont {R.~N.}\ \bibnamefont {Faustov}},
  \bibinfo {author} {\bibfnamefont {V.~O.}\ \bibnamefont {Galkin}}, \ and\
  \bibinfo {author} {\bibfnamefont {W.}~\bibnamefont {Lucha}},\ }\href
  {\doibase 10.1103/PhysRevD.76.114015} {\bibfield  {journal} {\bibinfo
  {journal} {Phys. Rev. D}\ }\textbf {\bibinfo {volume} {76}},\ \bibinfo
  {pages} {114015} (\bibinfo {year} {2007})},\ \Eprint
  {http://arxiv.org/abs/0706.3853} {arXiv:0706.3853 [hep-ph]} \BibitemShut
  {NoStop}%
\bibitem [{\citenamefont {Giron}\ and\ \citenamefont
  {Lebed}(2020)}]{Giron:2020wpx}%
  \BibitemOpen
  \bibfield  {author} {\bibinfo {author} {\bibfnamefont {J.~F.}\ \bibnamefont
  {Giron}}\ and\ \bibinfo {author} {\bibfnamefont {R.~F.}\ \bibnamefont
  {Lebed}},\ }\href {\doibase 10.1103/PhysRevD.102.074003} {\bibfield
  {journal} {\bibinfo  {journal} {Phys. Rev. D}\ }\textbf {\bibinfo {volume}
  {102}},\ \bibinfo {pages} {074003} (\bibinfo {year} {2020})},\ \Eprint
  {http://arxiv.org/abs/2008.01631} {arXiv:2008.01631 [hep-ph]} \BibitemShut
  {NoStop}%
\bibitem [{\citenamefont {Yang}\ \emph {et~al.}(2021)\citenamefont {Yang},
  \citenamefont {Ping},\ and\ \citenamefont {Segovia}}]{Yang:2021hrb}%
  \BibitemOpen
  \bibfield  {author} {\bibinfo {author} {\bibfnamefont {G.}~\bibnamefont
  {Yang}}, \bibinfo {author} {\bibfnamefont {J.}~\bibnamefont {Ping}}, \ and\
  \bibinfo {author} {\bibfnamefont {J.}~\bibnamefont {Segovia}},\ }\href
  {\doibase 10.1103/PhysRevD.104.014006} {\bibfield  {journal} {\bibinfo
  {journal} {Phys. Rev. D}\ }\textbf {\bibinfo {volume} {104}},\ \bibinfo
  {pages} {014006} (\bibinfo {year} {2021})},\ \Eprint
  {http://arxiv.org/abs/2104.08814} {arXiv:2104.08814 [hep-ph]} \BibitemShut
  {NoStop}%
\bibitem [{\citenamefont {Prelovsek}\ and\ \citenamefont
  {Leskovec}(2013)}]{Prelovsek:2013cra}%
  \BibitemOpen
  \bibfield  {author} {\bibinfo {author} {\bibfnamefont {S.}~\bibnamefont
  {Prelovsek}}\ and\ \bibinfo {author} {\bibfnamefont {L.}~\bibnamefont
  {Leskovec}},\ }\href {\doibase 10.1103/PhysRevLett.111.192001} {\bibfield
  {journal} {\bibinfo  {journal} {Phys. Rev. Lett.}\ }\textbf {\bibinfo
  {volume} {111}},\ \bibinfo {pages} {192001} (\bibinfo {year} {2013})},\
  \Eprint {http://arxiv.org/abs/1307.5172} {arXiv:1307.5172 [hep-lat]}
  \BibitemShut {NoStop}%
\bibitem [{\citenamefont {Ikeda}\ \emph {et~al.}(2014)\citenamefont {Ikeda},
  \citenamefont {Charron}, \citenamefont {Aoki}, \citenamefont {Doi},
  \citenamefont {Hatsuda}, \citenamefont {Inoue}, \citenamefont {Ishii},
  \citenamefont {Murano}, \citenamefont {Nemura},\ and\ \citenamefont
  {Sasaki}}]{Ikeda:2013vwa}%
  \BibitemOpen
  \bibfield  {author} {\bibinfo {author} {\bibfnamefont {Y.}~\bibnamefont
  {Ikeda}}, \bibinfo {author} {\bibfnamefont {B.}~\bibnamefont {Charron}},
  \bibinfo {author} {\bibfnamefont {S.}~\bibnamefont {Aoki}}, \bibinfo {author}
  {\bibfnamefont {T.}~\bibnamefont {Doi}}, \bibinfo {author} {\bibfnamefont
  {T.}~\bibnamefont {Hatsuda}}, \bibinfo {author} {\bibfnamefont
  {T.}~\bibnamefont {Inoue}}, \bibinfo {author} {\bibfnamefont
  {N.}~\bibnamefont {Ishii}}, \bibinfo {author} {\bibfnamefont
  {K.}~\bibnamefont {Murano}}, \bibinfo {author} {\bibfnamefont
  {H.}~\bibnamefont {Nemura}}, \ and\ \bibinfo {author} {\bibfnamefont
  {K.}~\bibnamefont {Sasaki}},\ }\href {\doibase
  10.1016/j.physletb.2014.01.002} {\bibfield  {journal} {\bibinfo  {journal}
  {Phys. Lett. B}\ }\textbf {\bibinfo {volume} {729}},\ \bibinfo {pages} {85}
  (\bibinfo {year} {2014})},\ \Eprint {http://arxiv.org/abs/1311.6214}
  {arXiv:1311.6214 [hep-lat]} \BibitemShut {NoStop}%
\bibitem [{\citenamefont {Prelovsek}\ \emph {et~al.}(2015)\citenamefont
  {Prelovsek}, \citenamefont {Lang}, \citenamefont {Leskovec},\ and\
  \citenamefont {Mohler}}]{Prelovsek:2014swa}%
  \BibitemOpen
  \bibfield  {author} {\bibinfo {author} {\bibfnamefont {S.}~\bibnamefont
  {Prelovsek}}, \bibinfo {author} {\bibfnamefont {C.~B.}\ \bibnamefont {Lang}},
  \bibinfo {author} {\bibfnamefont {L.}~\bibnamefont {Leskovec}}, \ and\
  \bibinfo {author} {\bibfnamefont {D.}~\bibnamefont {Mohler}},\ }\href
  {\doibase 10.1103/PhysRevD.91.014504} {\bibfield  {journal} {\bibinfo
  {journal} {Phys. Rev.}\ }\textbf {\bibinfo {volume} {D91}},\ \bibinfo {pages}
  {014504} (\bibinfo {year} {2015})},\ \Eprint {http://arxiv.org/abs/1405.7623}
  {arXiv:1405.7623 [hep-lat]} \BibitemShut {NoStop}%
\bibitem [{\citenamefont {Padmanath}\ \emph {et~al.}(2015)\citenamefont
  {Padmanath}, \citenamefont {Lang},\ and\ \citenamefont
  {Prelovsek}}]{Padmanath:2015era}%
  \BibitemOpen
  \bibfield  {author} {\bibinfo {author} {\bibfnamefont {M.}~\bibnamefont
  {Padmanath}}, \bibinfo {author} {\bibfnamefont {C.~B.}\ \bibnamefont {Lang}},
  \ and\ \bibinfo {author} {\bibfnamefont {S.}~\bibnamefont {Prelovsek}},\
  }\href {\doibase 10.1103/PhysRevD.92.034501} {\bibfield  {journal} {\bibinfo
  {journal} {Phys. Rev.}\ }\textbf {\bibinfo {volume} {D92}},\ \bibinfo {pages}
  {034501} (\bibinfo {year} {2015})},\ \Eprint
  {http://arxiv.org/abs/1503.03257} {arXiv:1503.03257 [hep-lat]} \BibitemShut
  {NoStop}%
\bibitem [{\citenamefont {Francis}\ \emph {et~al.}(2017)\citenamefont
  {Francis}, \citenamefont {Hudspith}, \citenamefont {Lewis},\ and\
  \citenamefont {Maltman}}]{Francis:2016hui}%
  \BibitemOpen
  \bibfield  {author} {\bibinfo {author} {\bibfnamefont {A.}~\bibnamefont
  {Francis}}, \bibinfo {author} {\bibfnamefont {R.~J.}\ \bibnamefont
  {Hudspith}}, \bibinfo {author} {\bibfnamefont {R.}~\bibnamefont {Lewis}}, \
  and\ \bibinfo {author} {\bibfnamefont {K.}~\bibnamefont {Maltman}},\ }\href
  {\doibase 10.1103/PhysRevLett.118.142001} {\bibfield  {journal} {\bibinfo
  {journal} {Phys. Rev. Lett.}\ }\textbf {\bibinfo {volume} {118}},\ \bibinfo
  {pages} {142001} (\bibinfo {year} {2017})},\ \Eprint
  {http://arxiv.org/abs/1607.05214} {arXiv:1607.05214 [hep-lat]} \BibitemShut
  {NoStop}%
\bibitem [{\citenamefont {Bicudo}\ \emph {et~al.}(2017)\citenamefont {Bicudo},
  \citenamefont {Cardoso}, \citenamefont {Peters}, \citenamefont {Pflaumer},\
  and\ \citenamefont {Wagner}}]{Bicudo:2017szl}%
  \BibitemOpen
  \bibfield  {author} {\bibinfo {author} {\bibfnamefont {P.}~\bibnamefont
  {Bicudo}}, \bibinfo {author} {\bibfnamefont {M.}~\bibnamefont {Cardoso}},
  \bibinfo {author} {\bibfnamefont {A.}~\bibnamefont {Peters}}, \bibinfo
  {author} {\bibfnamefont {M.}~\bibnamefont {Pflaumer}}, \ and\ \bibinfo
  {author} {\bibfnamefont {M.}~\bibnamefont {Wagner}},\ }\href {\doibase
  10.1103/PhysRevD.96.054510} {\bibfield  {journal} {\bibinfo  {journal} {Phys.
  Rev.}\ }\textbf {\bibinfo {volume} {D96}},\ \bibinfo {pages} {054510}
  (\bibinfo {year} {2017})},\ \Eprint {http://arxiv.org/abs/1704.02383}
  {arXiv:1704.02383 [hep-lat]} \BibitemShut {NoStop}%
\bibitem [{\citenamefont {Cheung}\ \emph {et~al.}(2017)\citenamefont {Cheung},
  \citenamefont {Thomas}, \citenamefont {Dudek},\ and\ \citenamefont
  {Edwards}}]{Cheung:2017tnt}%
  \BibitemOpen
  \bibfield  {author} {\bibinfo {author} {\bibfnamefont {G.~K.~C.}\
  \bibnamefont {Cheung}}, \bibinfo {author} {\bibfnamefont {C.~E.}\
  \bibnamefont {Thomas}}, \bibinfo {author} {\bibfnamefont {J.~J.}\
  \bibnamefont {Dudek}}, \ and\ \bibinfo {author} {\bibfnamefont {R.~G.}\
  \bibnamefont {Edwards}} (\bibinfo {collaboration} {Hadron Spectrum}),\ }\href
  {\doibase 10.1007/JHEP11(2017)033} {\bibfield  {journal} {\bibinfo  {journal}
  {JHEP}\ }\textbf {\bibinfo {volume} {11}},\ \bibinfo {pages} {033} (\bibinfo
  {year} {2017})},\ \Eprint {http://arxiv.org/abs/1709.01417} {arXiv:1709.01417
  [hep-lat]} \BibitemShut {NoStop}%
\bibitem [{\citenamefont {Francis}\ \emph {et~al.}(2019)\citenamefont
  {Francis}, \citenamefont {Hudspith}, \citenamefont {Lewis},\ and\
  \citenamefont {Maltman}}]{Francis:2018jyb}%
  \BibitemOpen
  \bibfield  {author} {\bibinfo {author} {\bibfnamefont {A.}~\bibnamefont
  {Francis}}, \bibinfo {author} {\bibfnamefont {R.~J.}\ \bibnamefont
  {Hudspith}}, \bibinfo {author} {\bibfnamefont {R.}~\bibnamefont {Lewis}}, \
  and\ \bibinfo {author} {\bibfnamefont {K.}~\bibnamefont {Maltman}},\ }\href
  {\doibase 10.1103/PhysRevD.99.054505} {\bibfield  {journal} {\bibinfo
  {journal} {Phys. Rev.}\ }\textbf {\bibinfo {volume} {D99}},\ \bibinfo {pages}
  {054505} (\bibinfo {year} {2019})},\ \Eprint
  {http://arxiv.org/abs/1810.10550} {arXiv:1810.10550 [hep-lat]} \BibitemShut
  {NoStop}%
\bibitem [{\citenamefont {Junnarkar}\ \emph {et~al.}(2019)\citenamefont
  {Junnarkar}, \citenamefont {Mathur},\ and\ \citenamefont
  {Padmanath}}]{Junnarkar:2018twb}%
  \BibitemOpen
  \bibfield  {author} {\bibinfo {author} {\bibfnamefont {P.}~\bibnamefont
  {Junnarkar}}, \bibinfo {author} {\bibfnamefont {N.}~\bibnamefont {Mathur}}, \
  and\ \bibinfo {author} {\bibfnamefont {M.}~\bibnamefont {Padmanath}},\ }\href
  {\doibase 10.1103/PhysRevD.99.034507} {\bibfield  {journal} {\bibinfo
  {journal} {Phys. Rev. D}\ }\textbf {\bibinfo {volume} {99}},\ \bibinfo
  {pages} {034507} (\bibinfo {year} {2019})},\ \Eprint
  {http://arxiv.org/abs/1810.12285} {arXiv:1810.12285 [hep-lat]} \BibitemShut
  {NoStop}%
\bibitem [{\citenamefont {Leskovec}\ \emph {et~al.}(2019)\citenamefont
  {Leskovec}, \citenamefont {Meinel}, \citenamefont {Pflaumer},\ and\
  \citenamefont {Wagner}}]{Leskovec:2019ioa}%
  \BibitemOpen
  \bibfield  {author} {\bibinfo {author} {\bibfnamefont {L.}~\bibnamefont
  {Leskovec}}, \bibinfo {author} {\bibfnamefont {S.}~\bibnamefont {Meinel}},
  \bibinfo {author} {\bibfnamefont {M.}~\bibnamefont {Pflaumer}}, \ and\
  \bibinfo {author} {\bibfnamefont {M.}~\bibnamefont {Wagner}},\ }\href
  {\doibase 10.1103/PhysRevD.100.014503} {\bibfield  {journal} {\bibinfo
  {journal} {Phys. Rev.}\ }\textbf {\bibinfo {volume} {D100}},\ \bibinfo
  {pages} {014503} (\bibinfo {year} {2019})},\ \Eprint
  {http://arxiv.org/abs/1904.04197} {arXiv:1904.04197 [hep-lat]} \BibitemShut
  {NoStop}%
\bibitem [{\citenamefont {Albuquerque}\ \emph {et~al.}(2019)\citenamefont
  {Albuquerque}, \citenamefont {Dias}, \citenamefont {Khemchandani},
  \citenamefont {Mart\'\i{}nez~Torres}, \citenamefont {Navarra}, \citenamefont
  {Nielsen},\ and\ \citenamefont {Zanetti}}]{Albuquerque:2018jkn}%
  \BibitemOpen
  \bibfield  {author} {\bibinfo {author} {\bibfnamefont {R.~M.}\ \bibnamefont
  {Albuquerque}}, \bibinfo {author} {\bibfnamefont {J.~M.}\ \bibnamefont
  {Dias}}, \bibinfo {author} {\bibfnamefont {K.~P.}\ \bibnamefont
  {Khemchandani}}, \bibinfo {author} {\bibfnamefont {A.}~\bibnamefont
  {Mart\'\i{}nez~Torres}}, \bibinfo {author} {\bibfnamefont {F.~S.}\
  \bibnamefont {Navarra}}, \bibinfo {author} {\bibfnamefont {M.}~\bibnamefont
  {Nielsen}}, \ and\ \bibinfo {author} {\bibfnamefont {C.~M.}\ \bibnamefont
  {Zanetti}},\ }\href {\doibase 10.1088/1361-6471/ab2678} {\bibfield  {journal}
  {\bibinfo  {journal} {J. Phys. G}\ }\textbf {\bibinfo {volume} {46}},\
  \bibinfo {pages} {093002} (\bibinfo {year} {2019})},\ \Eprint
  {http://arxiv.org/abs/1812.08207} {arXiv:1812.08207 [hep-ph]} \BibitemShut
  {NoStop}%
\bibitem [{\citenamefont {Wang}\ \emph {et~al.}(2013)\citenamefont {Wang},
  \citenamefont {Hanhart},\ and\ \citenamefont {Zhao}}]{Wang:2013cya}%
  \BibitemOpen
  \bibfield  {author} {\bibinfo {author} {\bibfnamefont {Q.}~\bibnamefont
  {Wang}}, \bibinfo {author} {\bibfnamefont {C.}~\bibnamefont {Hanhart}}, \
  and\ \bibinfo {author} {\bibfnamefont {Q.}~\bibnamefont {Zhao}},\ }\href
  {\doibase 10.1103/PhysRevLett.111.132003} {\bibfield  {journal} {\bibinfo
  {journal} {Phys. Rev. Lett.}\ }\textbf {\bibinfo {volume} {111}},\ \bibinfo
  {pages} {132003} (\bibinfo {year} {2013})},\ \Eprint
  {http://arxiv.org/abs/1303.6355} {arXiv:1303.6355 [hep-ph]} \BibitemShut
  {NoStop}%
\bibitem [{\citenamefont {Baru}\ \emph {et~al.}(2022)\citenamefont {Baru},
  \citenamefont {Epelbaum}, \citenamefont {Filin}, \citenamefont {Hanhart},\
  and\ \citenamefont {Nefediev}}]{Baru:2021ddn}%
  \BibitemOpen
  \bibfield  {author} {\bibinfo {author} {\bibfnamefont {V.}~\bibnamefont
  {Baru}}, \bibinfo {author} {\bibfnamefont {E.}~\bibnamefont {Epelbaum}},
  \bibinfo {author} {\bibfnamefont {A.~A.}\ \bibnamefont {Filin}}, \bibinfo
  {author} {\bibfnamefont {C.}~\bibnamefont {Hanhart}}, \ and\ \bibinfo
  {author} {\bibfnamefont {A.~V.}\ \bibnamefont {Nefediev}},\ }\href {\doibase
  10.1103/PhysRevD.105.034014} {\bibfield  {journal} {\bibinfo  {journal}
  {Phys. Rev. D}\ }\textbf {\bibinfo {volume} {105}},\ \bibinfo {pages}
  {034014} (\bibinfo {year} {2022})},\ \Eprint
  {http://arxiv.org/abs/2110.00398} {arXiv:2110.00398 [hep-ph]} \BibitemShut
  {NoStop}%
\bibitem [{\citenamefont {Wallbott}\ \emph {et~al.}(2019)\citenamefont
  {Wallbott}, \citenamefont {Eichmann},\ and\ \citenamefont
  {Fischer}}]{Wallbott:2019dng}%
  \BibitemOpen
  \bibfield  {author} {\bibinfo {author} {\bibfnamefont {P.~C.}\ \bibnamefont
  {Wallbott}}, \bibinfo {author} {\bibfnamefont {G.}~\bibnamefont {Eichmann}},
  \ and\ \bibinfo {author} {\bibfnamefont {C.~S.}\ \bibnamefont {Fischer}},\
  }\href {\doibase 10.1103/PhysRevD.100.014033} {\bibfield  {journal} {\bibinfo
   {journal} {Phys. Rev. D}\ }\textbf {\bibinfo {volume} {100}},\ \bibinfo
  {pages} {014033} (\bibinfo {year} {2019})},\ \Eprint
  {http://arxiv.org/abs/1905.02615} {arXiv:1905.02615 [hep-ph]} \BibitemShut
  {NoStop}%
\bibitem [{\citenamefont {Eichmann}\ \emph {et~al.}(2020)\citenamefont
  {Eichmann}, \citenamefont {Fischer}, \citenamefont {Heupel}, \citenamefont
  {Santowsky},\ and\ \citenamefont {Wallbott}}]{Eichmann:2020oqt}%
  \BibitemOpen
  \bibfield  {author} {\bibinfo {author} {\bibfnamefont {G.}~\bibnamefont
  {Eichmann}}, \bibinfo {author} {\bibfnamefont {C.~S.}\ \bibnamefont
  {Fischer}}, \bibinfo {author} {\bibfnamefont {W.}~\bibnamefont {Heupel}},
  \bibinfo {author} {\bibfnamefont {N.}~\bibnamefont {Santowsky}}, \ and\
  \bibinfo {author} {\bibfnamefont {P.~C.}\ \bibnamefont {Wallbott}},\ }\href
  {\doibase 10.1007/s00601-020-01571-3} {\bibfield  {journal} {\bibinfo
  {journal} {Few Body Syst.}\ }\textbf {\bibinfo {volume} {61}},\ \bibinfo
  {pages} {38} (\bibinfo {year} {2020})},\ \Eprint
  {http://arxiv.org/abs/2008.10240} {arXiv:2008.10240 [hep-ph]} \BibitemShut
  {NoStop}%
\bibitem [{\citenamefont {Wallbott}\ \emph {et~al.}(2020)\citenamefont
  {Wallbott}, \citenamefont {Eichmann},\ and\ \citenamefont
  {Fischer}}]{Wallbott:2020jzh}%
  \BibitemOpen
  \bibfield  {author} {\bibinfo {author} {\bibfnamefont {P.~C.}\ \bibnamefont
  {Wallbott}}, \bibinfo {author} {\bibfnamefont {G.}~\bibnamefont {Eichmann}},
  \ and\ \bibinfo {author} {\bibfnamefont {C.~S.}\ \bibnamefont {Fischer}},\
  }\href {\doibase 10.1103/PhysRevD.102.051501} {\bibfield  {journal} {\bibinfo
   {journal} {Phys. Rev. D}\ }\textbf {\bibinfo {volume} {102}},\ \bibinfo
  {pages} {051501} (\bibinfo {year} {2020})},\ \Eprint
  {http://arxiv.org/abs/2003.12407} {arXiv:2003.12407 [hep-ph]} \BibitemShut
  {NoStop}%
\bibitem [{\citenamefont {Dubynskiy}\ and\ \citenamefont
  {Voloshin}(2008)}]{Dubynskiy:2008mq}%
  \BibitemOpen
  \bibfield  {author} {\bibinfo {author} {\bibfnamefont {S.}~\bibnamefont
  {Dubynskiy}}\ and\ \bibinfo {author} {\bibfnamefont {M.~B.}\ \bibnamefont
  {Voloshin}},\ }\href {\doibase 10.1016/j.physletb.2008.07.086} {\bibfield
  {journal} {\bibinfo  {journal} {Phys. Lett. B}\ }\textbf {\bibinfo {volume}
  {666}},\ \bibinfo {pages} {344} (\bibinfo {year} {2008})},\ \Eprint
  {http://arxiv.org/abs/0803.2224} {arXiv:0803.2224 [hep-ph]} \BibitemShut
  {NoStop}%
\bibitem [{\citenamefont {Jaffe}(2005)}]{Jaffe:2004ph}%
  \BibitemOpen
  \bibfield  {author} {\bibinfo {author} {\bibfnamefont {R.~L.}\ \bibnamefont
  {Jaffe}},\ }\href {\doibase 10.1016/j.physrep.2004.11.005} {\bibfield
  {journal} {\bibinfo  {journal} {Phys. Rept.}\ }\textbf {\bibinfo {volume}
  {409}},\ \bibinfo {pages} {1} (\bibinfo {year} {2005})},\ \Eprint
  {http://arxiv.org/abs/hep-ph/0409065} {arXiv:hep-ph/0409065 [hep-ph]}
  \BibitemShut {NoStop}%
\bibitem [{\citenamefont {Aaij}\ \emph
  {et~al.}(2021{\natexlab{b}})\citenamefont {Aaij} \emph
  {et~al.}}]{LHCb:2021auc}%
  \BibitemOpen
  \bibfield  {author} {\bibinfo {author} {\bibfnamefont {R.}~\bibnamefont
  {Aaij}} \emph {et~al.} (\bibinfo {collaboration} {LHCb}),\ }\href@noop {} {\
  (\bibinfo {year} {2021}{\natexlab{b}})},\ \Eprint
  {http://arxiv.org/abs/2109.01056} {arXiv:2109.01056 [hep-ex]} \BibitemShut
  {NoStop}%
\bibitem [{\citenamefont {Du}\ \emph {et~al.}(2022)\citenamefont {Du},
  \citenamefont {Baru}, \citenamefont {Dong}, \citenamefont {Filin},
  \citenamefont {Guo}, \citenamefont {Hanhart}, \citenamefont {Nefediev},
  \citenamefont {Nieves},\ and\ \citenamefont {Wang}}]{Du:2021zzh}%
  \BibitemOpen
  \bibfield  {author} {\bibinfo {author} {\bibfnamefont {M.-L.}\ \bibnamefont
  {Du}}, \bibinfo {author} {\bibfnamefont {V.}~\bibnamefont {Baru}}, \bibinfo
  {author} {\bibfnamefont {X.-K.}\ \bibnamefont {Dong}}, \bibinfo {author}
  {\bibfnamefont {A.}~\bibnamefont {Filin}}, \bibinfo {author} {\bibfnamefont
  {F.-K.}\ \bibnamefont {Guo}}, \bibinfo {author} {\bibfnamefont
  {C.}~\bibnamefont {Hanhart}}, \bibinfo {author} {\bibfnamefont
  {A.}~\bibnamefont {Nefediev}}, \bibinfo {author} {\bibfnamefont
  {J.}~\bibnamefont {Nieves}}, \ and\ \bibinfo {author} {\bibfnamefont
  {Q.}~\bibnamefont {Wang}},\ }\href {\doibase 10.1103/PhysRevD.105.014024}
  {\bibfield  {journal} {\bibinfo  {journal} {Phys. Rev. D}\ }\textbf {\bibinfo
  {volume} {105}},\ \bibinfo {pages} {014024} (\bibinfo {year} {2022})},\
  \Eprint {http://arxiv.org/abs/2110.13765} {arXiv:2110.13765 [hep-ph]}
  \BibitemShut {NoStop}%
\bibitem [{\citenamefont {Heupel}\ \emph {et~al.}(2012)\citenamefont {Heupel},
  \citenamefont {Eichmann},\ and\ \citenamefont {Fischer}}]{Heupel:2012ua}%
  \BibitemOpen
  \bibfield  {author} {\bibinfo {author} {\bibfnamefont {W.}~\bibnamefont
  {Heupel}}, \bibinfo {author} {\bibfnamefont {G.}~\bibnamefont {Eichmann}}, \
  and\ \bibinfo {author} {\bibfnamefont {C.~S.}\ \bibnamefont {Fischer}},\
  }\href {\doibase 10.1016/j.physletb.2012.11.009} {\bibfield  {journal}
  {\bibinfo  {journal} {Phys. Lett. B}\ }\textbf {\bibinfo {volume} {718}},\
  \bibinfo {pages} {545} (\bibinfo {year} {2012})},\ \Eprint
  {http://arxiv.org/abs/1206.5129} {arXiv:1206.5129 [hep-ph]} \BibitemShut
  {NoStop}%
\bibitem [{\citenamefont {Santowsky}\ and\ \citenamefont
  {Fischer}(2022)}]{Santowsky:2021ugd}%
  \BibitemOpen
  \bibfield  {author} {\bibinfo {author} {\bibfnamefont {N.}~\bibnamefont
  {Santowsky}}\ and\ \bibinfo {author} {\bibfnamefont {C.~S.}\ \bibnamefont
  {Fischer}},\ }\href {\doibase 10.1103/PhysRevD.105.034025} {\bibfield
  {journal} {\bibinfo  {journal} {Phys. Rev. D}\ }\textbf {\bibinfo {volume}
  {105}},\ \bibinfo {pages} {034025} (\bibinfo {year} {2022})},\ \Eprint
  {http://arxiv.org/abs/2109.00755} {arXiv:2109.00755 [hep-ph]} \BibitemShut
  {NoStop}%
\bibitem [{\citenamefont {Santowsky}\ \emph {et~al.}(2020)\citenamefont
  {Santowsky}, \citenamefont {Eichmann}, \citenamefont {Fischer}, \citenamefont
  {Wallbott},\ and\ \citenamefont {Williams}}]{Santowsky:2020pwd}%
  \BibitemOpen
  \bibfield  {author} {\bibinfo {author} {\bibfnamefont {N.}~\bibnamefont
  {Santowsky}}, \bibinfo {author} {\bibfnamefont {G.}~\bibnamefont {Eichmann}},
  \bibinfo {author} {\bibfnamefont {C.~S.}\ \bibnamefont {Fischer}}, \bibinfo
  {author} {\bibfnamefont {P.~C.}\ \bibnamefont {Wallbott}}, \ and\ \bibinfo
  {author} {\bibfnamefont {R.}~\bibnamefont {Williams}},\ }\href {\doibase
  10.1103/PhysRevD.102.056014} {\bibfield  {journal} {\bibinfo  {journal}
  {Phys. Rev. D}\ }\textbf {\bibinfo {volume} {102}},\ \bibinfo {pages}
  {056014} (\bibinfo {year} {2020})},\ \Eprint
  {http://arxiv.org/abs/2007.06495} {arXiv:2007.06495 [hep-ph]} \BibitemShut
  {NoStop}%
\bibitem [{\citenamefont {Kvinikhidze}\ and\ \citenamefont
  {Blankleider}(2014)}]{Kvinikhidze:2014yqa}%
  \BibitemOpen
  \bibfield  {author} {\bibinfo {author} {\bibfnamefont {A.}~\bibnamefont
  {Kvinikhidze}}\ and\ \bibinfo {author} {\bibfnamefont {B.}~\bibnamefont
  {Blankleider}},\ }\href {\doibase 10.1103/PhysRevD.90.045042} {\bibfield
  {journal} {\bibinfo  {journal} {Phys. Rev. D}\ }\textbf {\bibinfo {volume}
  {90}},\ \bibinfo {pages} {045042} (\bibinfo {year} {2014})},\ \Eprint
  {http://arxiv.org/abs/1406.5599} {arXiv:1406.5599 [hep-ph]} \BibitemShut
  {NoStop}%
\bibitem [{\citenamefont {Kvinikhidze}\ and\ \citenamefont
  {Blankleider}(2021)}]{Kvinikhidze:2021kzu}%
  \BibitemOpen
  \bibfield  {author} {\bibinfo {author} {\bibfnamefont {A.~N.}\ \bibnamefont
  {Kvinikhidze}}\ and\ \bibinfo {author} {\bibfnamefont {B.}~\bibnamefont
  {Blankleider}},\ }\href@noop {} {\  (\bibinfo {year} {2021})},\ \Eprint
  {http://arxiv.org/abs/2102.09558} {arXiv:2102.09558 [hep-th]} \BibitemShut
  {NoStop}%
\bibitem [{\citenamefont {Zyla}\ \emph {et~al.}(2020)\citenamefont {Zyla} \emph
  {et~al.}}]{ParticleDataGroup:2020ssz}%
  \BibitemOpen
  \bibfield  {author} {\bibinfo {author} {\bibfnamefont {P.~A.}\ \bibnamefont
  {Zyla}} \emph {et~al.} (\bibinfo {collaboration} {Particle Data Group}),\
  }\href {\doibase 10.1093/ptep/ptaa104} {\bibfield  {journal} {\bibinfo
  {journal} {PTEP}\ }\textbf {\bibinfo {volume} {2020}},\ \bibinfo {pages}
  {083C01} (\bibinfo {year} {2020})}\BibitemShut {NoStop}%
\bibitem [{\citenamefont {Eichmann}\ \emph
  {et~al.}(2016{\natexlab{a}})\citenamefont {Eichmann}, \citenamefont
  {Fischer},\ and\ \citenamefont {Sanchis-Alepuz}}]{Eichmann:2016hgl}%
  \BibitemOpen
  \bibfield  {author} {\bibinfo {author} {\bibfnamefont {G.}~\bibnamefont
  {Eichmann}}, \bibinfo {author} {\bibfnamefont {C.~S.}\ \bibnamefont
  {Fischer}}, \ and\ \bibinfo {author} {\bibfnamefont {H.}~\bibnamefont
  {Sanchis-Alepuz}},\ }\href {\doibase 10.1103/PhysRevD.94.094033} {\bibfield
  {journal} {\bibinfo  {journal} {Phys. Rev. D}\ }\textbf {\bibinfo {volume}
  {94}},\ \bibinfo {pages} {094033} (\bibinfo {year} {2016}{\natexlab{a}})},\
  \Eprint {http://arxiv.org/abs/1607.05748} {arXiv:1607.05748 [hep-ph]}
  \BibitemShut {NoStop}%
\bibitem [{\citenamefont {Eichmann}\ \emph
  {et~al.}(2016{\natexlab{b}})\citenamefont {Eichmann}, \citenamefont
  {Sanchis-Alepuz}, \citenamefont {Williams}, \citenamefont {Alkofer},\ and\
  \citenamefont {Fischer}}]{Eichmann:2016yit}%
  \BibitemOpen
  \bibfield  {author} {\bibinfo {author} {\bibfnamefont {G.}~\bibnamefont
  {Eichmann}}, \bibinfo {author} {\bibfnamefont {H.}~\bibnamefont
  {Sanchis-Alepuz}}, \bibinfo {author} {\bibfnamefont {R.}~\bibnamefont
  {Williams}}, \bibinfo {author} {\bibfnamefont {R.}~\bibnamefont {Alkofer}}, \
  and\ \bibinfo {author} {\bibfnamefont {C.~S.}\ \bibnamefont {Fischer}},\
  }\href {\doibase 10.1016/j.ppnp.2016.07.001} {\bibfield  {journal} {\bibinfo
  {journal} {Prog. Part. Nucl. Phys.}\ }\textbf {\bibinfo {volume} {91}},\
  \bibinfo {pages} {1} (\bibinfo {year} {2016}{\natexlab{b}})},\ \Eprint
  {http://arxiv.org/abs/1606.09602} {arXiv:1606.09602 [hep-ph]} \BibitemShut
  {NoStop}%
\bibitem [{\citenamefont {Sanchis-Alepuz}\ and\ \citenamefont
  {Williams}(2018)}]{Sanchis-Alepuz:2017jjd}%
  \BibitemOpen
  \bibfield  {author} {\bibinfo {author} {\bibfnamefont {H.}~\bibnamefont
  {Sanchis-Alepuz}}\ and\ \bibinfo {author} {\bibfnamefont {R.}~\bibnamefont
  {Williams}},\ }\href {\doibase 10.1016/j.cpc.2018.05.020} {\bibfield
  {journal} {\bibinfo  {journal} {Comput. Phys. Commun.}\ }\textbf {\bibinfo
  {volume} {232}},\ \bibinfo {pages} {1} (\bibinfo {year} {2018})},\ \Eprint
  {http://arxiv.org/abs/1710.04903} {arXiv:1710.04903 [hep-ph]} \BibitemShut
  {NoStop}%
\bibitem [{\citenamefont {Maris}\ and\ \citenamefont
  {Tandy}(1999)}]{Maris:1999nt}%
  \BibitemOpen
  \bibfield  {author} {\bibinfo {author} {\bibfnamefont {P.}~\bibnamefont
  {Maris}}\ and\ \bibinfo {author} {\bibfnamefont {P.~C.}\ \bibnamefont
  {Tandy}},\ }\href {\doibase 10.1103/PhysRevC.60.055214} {\bibfield  {journal}
  {\bibinfo  {journal} {Phys. Rev. C}\ }\textbf {\bibinfo {volume} {60}},\
  \bibinfo {pages} {055214} (\bibinfo {year} {1999})},\ \Eprint
  {http://arxiv.org/abs/nucl-th/9905056} {arXiv:nucl-th/9905056} \BibitemShut
  {NoStop}%
\bibitem [{\citenamefont {Williams}(2019)}]{Williams:2018adr}%
  \BibitemOpen
  \bibfield  {author} {\bibinfo {author} {\bibfnamefont {R.}~\bibnamefont
  {Williams}},\ }\href {\doibase 10.1016/j.physletb.2019.134943} {\bibfield
  {journal} {\bibinfo  {journal} {Phys. Lett. B}\ }\textbf {\bibinfo {volume}
  {798}},\ \bibinfo {pages} {134943} (\bibinfo {year} {2019})},\ \Eprint
  {http://arxiv.org/abs/1804.11161} {arXiv:1804.11161 [hep-ph]} \BibitemShut
  {NoStop}%
\bibitem [{\citenamefont {Santowsky}(2021)}]{Santowsky:10.22029}%
  \BibitemOpen
  \bibfield  {author} {\bibinfo {author} {\bibfnamefont {N.}~\bibnamefont
  {Santowsky}},\ }\emph {\bibinfo {title} {The Role of Four-Quark States in the
  Nature of Exotic Hadrons from bethe-Salpeter Equations}},\ \href@noop {}
  {Ph.D. thesis},\ \bibinfo  {school} {University of Giessen, Germany}
  (\bibinfo {year} {2021})\BibitemShut {NoStop}%
\bibitem [{\citenamefont {Windisch}(2017)}]{Windisch:2016iud}%
  \BibitemOpen
  \bibfield  {author} {\bibinfo {author} {\bibfnamefont {A.}~\bibnamefont
  {Windisch}},\ }\href {\doibase 10.1103/PhysRevC.95.045204} {\bibfield
  {journal} {\bibinfo  {journal} {Phys. Rev. C}\ }\textbf {\bibinfo {volume}
  {95}},\ \bibinfo {pages} {045204} (\bibinfo {year} {2017})},\ \Eprint
  {http://arxiv.org/abs/1612.06002} {arXiv:1612.06002 [hep-ph]} \BibitemShut
  {NoStop}%
\bibitem [{\citenamefont {Williams}\ \emph {et~al.}(2016)\citenamefont
  {Williams}, \citenamefont {Fischer},\ and\ \citenamefont
  {Heupel}}]{Williams:2015cvx}%
  \BibitemOpen
  \bibfield  {author} {\bibinfo {author} {\bibfnamefont {R.}~\bibnamefont
  {Williams}}, \bibinfo {author} {\bibfnamefont {C.~S.}\ \bibnamefont
  {Fischer}}, \ and\ \bibinfo {author} {\bibfnamefont {W.}~\bibnamefont
  {Heupel}},\ }\href {\doibase 10.1103/PhysRevD.93.034026} {\bibfield
  {journal} {\bibinfo  {journal} {Phys. Rev. D}\ }\textbf {\bibinfo {volume}
  {93}},\ \bibinfo {pages} {034026} (\bibinfo {year} {2016})},\ \Eprint
  {http://arxiv.org/abs/1512.00455} {arXiv:1512.00455 [hep-ph]} \BibitemShut
  {NoStop}%
\bibitem [{\citenamefont {Baru}\ \emph {et~al.}(2011)\citenamefont {Baru},
  \citenamefont {Filin}, \citenamefont {Hanhart}, \citenamefont {Kalashnikova},
  \citenamefont {Kudryavtsev},\ and\ \citenamefont {Nefediev}}]{Baru:2011rs}%
  \BibitemOpen
  \bibfield  {author} {\bibinfo {author} {\bibfnamefont {V.}~\bibnamefont
  {Baru}}, \bibinfo {author} {\bibfnamefont {A.~A.}\ \bibnamefont {Filin}},
  \bibinfo {author} {\bibfnamefont {C.}~\bibnamefont {Hanhart}}, \bibinfo
  {author} {\bibfnamefont {Y.~S.}\ \bibnamefont {Kalashnikova}}, \bibinfo
  {author} {\bibfnamefont {A.~E.}\ \bibnamefont {Kudryavtsev}}, \ and\ \bibinfo
  {author} {\bibfnamefont {A.~V.}\ \bibnamefont {Nefediev}},\ }\href {\doibase
  10.1103/PhysRevD.84.074029} {\bibfield  {journal} {\bibinfo  {journal} {Phys.
  Rev. D}\ }\textbf {\bibinfo {volume} {84}},\ \bibinfo {pages} {074029}
  (\bibinfo {year} {2011})},\ \Eprint {http://arxiv.org/abs/1108.5644}
  {arXiv:1108.5644 [hep-ph]} \BibitemShut {NoStop}%
\bibitem [{\citenamefont {Cleven}\ \emph {et~al.}(2015)\citenamefont {Cleven},
  \citenamefont {Guo}, \citenamefont {Hanhart}, \citenamefont {Wang},\ and\
  \citenamefont {Zhao}}]{Cleven:2015era}%
  \BibitemOpen
  \bibfield  {author} {\bibinfo {author} {\bibfnamefont {M.}~\bibnamefont
  {Cleven}}, \bibinfo {author} {\bibfnamefont {F.-K.}\ \bibnamefont {Guo}},
  \bibinfo {author} {\bibfnamefont {C.}~\bibnamefont {Hanhart}}, \bibinfo
  {author} {\bibfnamefont {Q.}~\bibnamefont {Wang}}, \ and\ \bibinfo {author}
  {\bibfnamefont {Q.}~\bibnamefont {Zhao}},\ }\href {\doibase
  10.1103/PhysRevD.92.014005} {\bibfield  {journal} {\bibinfo  {journal} {Phys.
  Rev.}\ }\textbf {\bibinfo {volume} {D92}},\ \bibinfo {pages} {014005}
  (\bibinfo {year} {2015})},\ \Eprint {http://arxiv.org/abs/1505.01771}
  {arXiv:1505.01771 [hep-ph]} \BibitemShut {NoStop}%
\bibitem [{\citenamefont {Manohar}\ and\ \citenamefont
  {Wise}(1993)}]{Manohar:1992nd}%
  \BibitemOpen
  \bibfield  {author} {\bibinfo {author} {\bibfnamefont {A.~V.}\ \bibnamefont
  {Manohar}}\ and\ \bibinfo {author} {\bibfnamefont {M.~B.}\ \bibnamefont
  {Wise}},\ }\href {\doibase 10.1016/0550-3213(93)90614-U} {\bibfield
  {journal} {\bibinfo  {journal} {Nucl. Phys. B}\ }\textbf {\bibinfo {volume}
  {399}},\ \bibinfo {pages} {17} (\bibinfo {year} {1993})},\ \Eprint
  {http://arxiv.org/abs/hep-ph/9212236} {arXiv:hep-ph/9212236} \BibitemShut
  {NoStop}%
\bibitem [{\citenamefont {Faustov}\ \emph {et~al.}(2020)\citenamefont
  {Faustov}, \citenamefont {Galkin},\ and\ \citenamefont
  {Savchenko}}]{Faustov:2020qfm}%
  \BibitemOpen
  \bibfield  {author} {\bibinfo {author} {\bibfnamefont {R.~N.}\ \bibnamefont
  {Faustov}}, \bibinfo {author} {\bibfnamefont {V.~O.}\ \bibnamefont {Galkin}},
  \ and\ \bibinfo {author} {\bibfnamefont {E.~M.}\ \bibnamefont {Savchenko}},\
  }\href {\doibase 10.1103/PhysRevD.102.114030} {\bibfield  {journal} {\bibinfo
   {journal} {Phys. Rev. D}\ }\textbf {\bibinfo {volume} {102}},\ \bibinfo
  {pages} {114030} (\bibinfo {year} {2020})},\ \Eprint
  {http://arxiv.org/abs/2009.13237} {arXiv:2009.13237 [hep-ph]} \BibitemShut
  {NoStop}%
\bibitem [{\citenamefont {Faustov}\ \emph {et~al.}(2021)\citenamefont
  {Faustov}, \citenamefont {Galkin},\ and\ \citenamefont
  {Savchenko}}]{Faustov:2021hjs}%
  \BibitemOpen
  \bibfield  {author} {\bibinfo {author} {\bibfnamefont {R.~N.}\ \bibnamefont
  {Faustov}}, \bibinfo {author} {\bibfnamefont {V.~O.}\ \bibnamefont {Galkin}},
  \ and\ \bibinfo {author} {\bibfnamefont {E.~M.}\ \bibnamefont {Savchenko}},\
  }\href {\doibase 10.3390/universe7040094} {\bibfield  {journal} {\bibinfo
  {journal} {Universe}\ }\textbf {\bibinfo {volume} {7}},\ \bibinfo {pages}
  {94} (\bibinfo {year} {2021})},\ \Eprint {http://arxiv.org/abs/2103.01763}
  {arXiv:2103.01763 [hep-ph]} \BibitemShut {NoStop}%
\bibitem [{\citenamefont {Zhao}\ \emph {et~al.}(2020)\citenamefont {Zhao},
  \citenamefont {Shi},\ and\ \citenamefont {Zhuang}}]{Zhao:2020nwy}%
  \BibitemOpen
  \bibfield  {author} {\bibinfo {author} {\bibfnamefont {J.}~\bibnamefont
  {Zhao}}, \bibinfo {author} {\bibfnamefont {S.}~\bibnamefont {Shi}}, \ and\
  \bibinfo {author} {\bibfnamefont {P.}~\bibnamefont {Zhuang}},\ }\href
  {\doibase 10.1103/PhysRevD.102.114001} {\bibfield  {journal} {\bibinfo
  {journal} {Phys. Rev. D}\ }\textbf {\bibinfo {volume} {102}},\ \bibinfo
  {pages} {114001} (\bibinfo {year} {2020})},\ \Eprint
  {http://arxiv.org/abs/2009.10319} {arXiv:2009.10319 [hep-ph]} \BibitemShut
  {NoStop}%
\bibitem [{\citenamefont {Deng}\ \emph {et~al.}(2021)\citenamefont {Deng},
  \citenamefont {Chen},\ and\ \citenamefont {Ping}}]{Deng:2020iqw}%
  \BibitemOpen
  \bibfield  {author} {\bibinfo {author} {\bibfnamefont {C.}~\bibnamefont
  {Deng}}, \bibinfo {author} {\bibfnamefont {H.}~\bibnamefont {Chen}}, \ and\
  \bibinfo {author} {\bibfnamefont {J.}~\bibnamefont {Ping}},\ }\href {\doibase
  10.1103/PhysRevD.103.014001} {\bibfield  {journal} {\bibinfo  {journal}
  {Phys. Rev. D}\ }\textbf {\bibinfo {volume} {103}},\ \bibinfo {pages}
  {014001} (\bibinfo {year} {2021})},\ \Eprint
  {http://arxiv.org/abs/2003.05154} {arXiv:2003.05154 [hep-ph]} \BibitemShut
  {NoStop}%
\end{thebibliography}%

\end{document}